\documentclass{article}
\usepackage{setspace}
\usepackage[utf8]{inputenc}
\usepackage[english]{babel}
\usepackage{amsfonts,amsmath,amssymb,mathtools}
\usepackage[left=2cm,right=2cm,top=2cm,bottom=2cm,bindingoffset=0cm]{geometry}
\usepackage{graphicx}
\usepackage{authblk}
\usepackage{MnSymbol}
\usepackage[pdftex]{pict2e}
\usepackage{xcolor}
\usepackage{hyperref}
\usepackage{dsfont}
\usepackage{authblk}
\usepackage{slashed}
\usepackage{braket}
\usepackage{comment}
\usepackage{pb-diagram}
\usepackage{wasysym}
\usepackage{amsthm}
\usepackage{cancel}
\usepackage{tikz}\usetikzlibrary{tikzmark}
\usetikzlibrary{fadings} 
\usetikzlibrary{positioning}
\usetikzlibrary{backgrounds} 
\usetikzlibrary{decorations.pathreplacing}
\usetikzlibrary{arrows}
\definecolor{airforceblue}{rgb}{0.36, 0.54, 0.66}	
\definecolor{beige}{rgb}{0.96, 0.96, 0.86}
\definecolor{bittersweet}{rgb}{1.0, 0.44, 0.37}
\definecolor{melon}{rgb}{0.99, 0.74, 0.71}
\definecolor{mustard}{rgb}{1.0, 0.86, 0.35}
\definecolor{lava}{rgb}{0.81, 0.06, 0.13}
\definecolor{magnolia}{rgb}{0.97, 0.96, 1.0}
\definecolor{lavendermist}{rgb}{0.9, 0.9, 0.98}
\definecolor{lavendergray}{rgb}{0.77, 0.76, 0.82}
\definecolor{palepink}{rgb}{0.98, 0.85, 0.87}
\definecolor{palesilver}{rgb}{0.79, 0.75, 0.73}
\definecolor{cadetgrey}{rgb}{0.57, 0.64, 0.69}
\definecolor{anti-flashwhite}{rgb}{0.95, 0.95, 0.96}
\colorlet{Light0anti-flashwhite}{anti-flashwhite!70!white}
\colorlet{Lightanti-flashwhite}{anti-flashwhite!50!white}
\colorlet{Light2anti-flashwhite}{anti-flashwhite!30!white}
\definecolor{linkcolor}{rgb}{0,0,1}
\definecolor{urlcolor}{rgb}{0,0,1}

\usepackage{tikz-cd}
\usetikzlibrary{decorations.markings}
\usetikzlibrary{positioning}
\usepackage{braids}
\usepackage{array}
\usepackage{ytableau}
\usepackage{longtable}
\usepackage{empheq}
\usepackage{multicol}
\usepackage{mathrsfs}
\usepackage{diagbox}
\usepackage[vcentermath]{youngtab}
\usepackage[natbib=true, backend = bibtex, style=numeric-comp, sorting=none,maxbibnames=99]{biblatex}
\hypersetup{pdfstartview=FitH,  linkcolor=linkcolor,urlcolor=urlcolor,citecolor=urlcolor, colorlinks=true}

\newcommand\bem{\begin{pmatrix}}
\newcommand\eem{\end{pmatrix}}
\newcommand\beq{\begin{equation}}
\newcommand\eeq{\end{equation}}
\newcommand\beqs{\begin{equation*}}
\newcommand\eeqs{\end{equation*}}

\date{}

\def\be{\begin{eqnarray}}
\def\ee{\end{eqnarray}}

\def\ba{\begin{equation}\begin{aligned}}
\def\ea{\end{aligned}\end{equation}}

\definecolor{red}{rgb}{1,0,0}
\definecolor{orange}{rgb}{1,0.5,0}
\definecolor{violet}{rgb}{0.7,0,1}

\newtheorem{definition}{Definition}[section]

\newtheorem{statement}{Statement}

\begin{document}

\title{\Large \bf Algebraic structures of Vassiliev invariants for knot families
}

\author[1,2,3]{{\bf E. Lanina}\thanks{\href{mailto:lanina.en@phystech.edu}{lanina.en@phystech.edu}}}
\author[1,2,3]{{\bf A. Sleptsov}\thanks{\href{mailto:sleptsov@itep.ru}{sleptsov@itep.ru}}}

\vspace{5cm}

\affil[1]{Moscow Institute of Physics and Technology, 141700, Dolgoprudny, Russia}
\affil[2]{Institute for Information Transmission Problems, 127051, Moscow, Russia}
\affil[3]{NRC "Kurchatov Institute", 123182, Moscow, Russia\footnote{former Institute for Theoretical and Experimental Physics, 117218, Moscow, Russia}}
\renewcommand\Affilfont{\itshape\small}

\maketitle

\vspace{-7cm}

\begin{center}
	\hfill MIPT/TH-19/25\\
	\hfill ITEP/TH-26/25\\
	\hfill IITP/TH-23/25
\end{center}

\vspace{4.5cm}

\begin{abstract}

{
We explore algebraic relations on Vassiliev knot invariants related with correlators in the 3-dimensional Chern--Simons theory. Vassiliev invariants form infinite-dimensional algebra. We focus on $k$-parametric knot families with Vassiliev invariants being polynomials in family parameters. We conjecture that such 1-parametric algebra of Vassiliev invariants is always finitely generated, while in the case of more parameters, we provide example of the knot family with infinite number of generators. Inside a knot family, there appear extra algebraic relations on Vassiliev invariants. We show that there are $\leq k$ algebraically independent Vassiliev invariants for $k$-parametric knot family. However, in all our examples, the number of algebraically independent Vassiliev invariants is exactly $k$, and it is open question if there exists a $k$-parametric knot family with a fewer number of algebraically independent Vassiliev invariants. We also demonstrate that a complete knot invariant of some $k$-parametric knot families consists of $k$ Vassiliev invariants. 
}
\end{abstract}

\tableofcontents


\section{Introduction}

We explore observables in the three-dimensional topological Chern--Simons theory~\cite{Chern--Simons,Schwarz} in $S^3$ with some simple compact Lie gauge group $G$ defined by the following action:
\begin{equation}
         S_{\text{CS}}[{\cal A}] = \frac{\kappa}{4 \pi} \int_{S^3} \text{tr} \left( {\cal A} \wedge d{\cal A} +  \frac{2}{3} {\cal A} \wedge {\cal A} \wedge {\cal A} \right).
\end{equation}
Correlators in this theory can be obtained from the gauge invariant Wilson loop by the expansion of the path ordered exponent:
\begin{equation}
     \label{WilsonLoopExpValue}
\begin{aligned}
          \frac{1}{\dim(R)}\left\langle \text{tr}_{R} \ \text{Pexp} \left( \oint_{\mathcal{K}} {\cal A} \right) \right\rangle_{\text{CS}} \, = \phantom{\hspace{4.5cm}}  \\ = \ 1 \ + \ \frac{\text{tr}_R\,(T^a)}{\dim(R)} \oint_{\mathcal{K}} \left\langle {\cal A}_i^a(x) \right\rangle \,dx^i \ + \ \frac{1}{2!}\,\frac{\text{tr}_R\, (T^{a_1}T^{a_2})}{\dim(R)}\,\oint_{\mathcal{K}} dx_2^i \int_0^{x_2}dx_1^i \left\langle {\cal A}_{i_1}^{a_1}(x_1)\,{\cal A}_{i_2}^{a_2}(x_2)\right\rangle \ + \ \ldots
\end{aligned}
\end{equation}
where the gauge fields are taken in an arbitrary representation ${\cal A}={\cal A}_\mu^a T_a dx^\mu$ with $T_a$ being generators of the corresponding Lie algebra ${\mathfrak g}$ in the representation $R$, and the integration contour can be tied in an arbitrary knot. Note that the knot and the group dependences split in each term of the expansion. Choosing a proper framing and normalisation factors (for details, see~\cite{Witten}), Wilson loop turns out to be a polynomial in special variables $q$ and $A$ depending on the coupling constant $\kappa$ and on the Lie algebra rank. We rewrite the Wilson loop expansion as
\begin{equation}
\begin{aligned}
     \label{LoopExpansionHOMFLY}
         &\mathcal{H}_{R}^{\mathcal{K}}(q,A) = \sum_{d=0}^{\infty} \hbar^{d} \sum_{i=1}^{\dim\left(\mathbb{G}_{d}^{{\mathfrak g}}\right)} \mathcal{V}^{\mathcal{K}}_{d,i}\,\mathcal{G}_{d, i}^R \, = \, 1 \ + \ \hbar^2 \mathcal{V}^{\mathcal{K}}_{2,1}\mathcal{G}_{2,1}^R \ + \ \hbar^3\mathcal{V}^{\mathcal{K}}_{3,1}\mathcal{G}_{3,1}^R \ + \ \hbar^4\left(\mathcal{V}^{\mathcal{K}}_{4,1}\mathcal{G}_{4,1}^R+\mathcal{V}^{\mathcal{K}}_{4,2}\mathcal{G}_{4,2}^R+\mathcal{V}^{\mathcal{K}}_{4,3}\mathcal{G}_{4,3}^R\right) \ + \ \dots
    \end{aligned}
     \end{equation}
where 
\begin{equation}\label{V-inv}
\begin{aligned}
    \mathcal{V}^{\mathcal{K}}_{d, i} \sim \oint d x_{1} \ldots \int d x_{2n}\left\langle {\cal A}^{i^{(i)}_{1}}\left(x_{1}\right) \ldots {\cal A}^{i^{(i)}_{2n}}\left(x_{2n}\right)\right\rangle 
\end{aligned}
\end{equation}
and
\begin{equation}\label{G-factors}
    \mathcal{G}_{d, i}^R \sim \text{tr}_{R} \left(T_{i^{(i)}_1} T_{i^{(i)}_2}\ldots T_{i^{(i)}_{2n}} \right)\,.
\end{equation}
In~\eqref{LoopExpansionHOMFLY}, $\dim\left(\mathbb{G}_{d}^{\mathfrak g}\right)$ is the number of linear independent {\it group factors} of order $d$ for a Lie algebra ${\mathfrak g}$. Since the Chern--Simons theory is topological, both the Wilson loop, and the knot-dependent summands in its expansion are knot invariants. In knot theory, functions $\mathcal{V}^{\mathcal{K}}_{d, i}$,~\eqref{V-inv}, are called {\it Vassiliev invariants}~\cite{Vassiliev} of order $\leq d$ and the representation-dependent multipliers $\mathcal{G}_{d, i}^R$,~\eqref{G-factors}, are called {\it group factors}~\cite{alvarez1995numerical,lanina2021chern,lanina2022implications}. For $SU(N)$ gauge group, this properly normalised Wilson loop~\eqref{LoopExpansionHOMFLY} is called the colored HOMFLY polynomial~\cite{freyd1985new,przytycki1987kobe}; its $q$ and $A$ variables are related to Chern--Simons parameters
   as follows: $q=\exp\left(\frac{2 \pi i}{\kappa + N}\right)$, $A=q^N$.

Behind the perturbative expansion~\eqref{LoopExpansionHOMFLY}, there stands the mathematical construction called {\it Kontsevich integral} or the {\it universal Vassiliev invariant}~\cite{kontsevich1993vassiliev,chmutov2012introduction}\,:
\begin{equation}
    I^{\cal K} = \sum\limits_{d=0}^{\infty} \hbar^d \sum\limits_{i=1}^{\dim({\cal A}_d)} \mathcal{V}^{\mathcal{K}}_{d,i}\,D_{d, i}
\end{equation}
where $D_{d, i}$ are chord diagrams~\cite{chmutov2012introduction}. Each degree $d$ component of universal Vassiliev
   invariant contains all $\dim({\cal A}_d)$ Vassiliev invariants of order $\leq d$. One can descend from Kontsevich integral to the perturbative expansion of the Wilson loop~\eqref{LoopExpansionHOMFLY} with the use of the linear map $\varphi_{{\mathfrak g}}^R$ called a Lie algebra ${\mathfrak g}$ weight system~\cite{chmutov2012introduction,khudoteplov2024construction} associated with a representation $R$:
\begin{equation}
    \varphi_{{\mathfrak g}}^R(D_{d,i}) = \mathcal{G}_{d, i}^R \quad \Longrightarrow \quad \varphi_{{\mathfrak g}}^R(I^{\cal K}) = {\cal H}_R^{\cal K}(q,A)\,.
\end{equation}
Vogel has proved~\cite{vogel1999universal,vogel2011algebraic,khudoteplov2024construction,khudoteplov2025can,morozov2025vogel} that for any (semi)simple (super)algebra Lie, ${\rm Ker}\, \varphi_{{\mathfrak g}}^R \neq \varnothing\,$, i.e. $\dim({\cal A}_d) \geq\, \dim(\mathbb{G}_d^{\mathfrak g})\;\forall\,{\mathfrak g}\,$. The numbers of Vassiliev invariants and group factors for the Lie algebra $\mathfrak{g}=\mathfrak{sl}_N$ at orders $d=1,\dots,13$ are provided in Table~\ref{dimV&G}.

\begin{table}[h!]\label{dimV&G}
\begin{center}
\begin{tabular}{|c||c|c|c|c|c|c|c|c|c|c|c|c|c|}
    \hline
    \raisebox{-0.1cm}{$d$} & \raisebox{-0.1cm}{1} & \raisebox{-0.1cm}{2} & \raisebox{-0.1cm}{3} & \raisebox{-0.1cm}{4} & \raisebox{-0.1cm}{5} & \raisebox{-0.1cm}{6} & \raisebox{-0.1cm}{7} & \raisebox{-0.1cm}{8} & \raisebox{-0.1cm}{9} & \raisebox{-0.1cm}{10} & \raisebox{-0.1cm}{11} & \raisebox{-0.1cm}{12} & \raisebox{-0.1cm}{13} \\ [1.5ex]
    \hline
    \hline
    \raisebox{-0.1cm}{$\dim \mathcal{A}_d$} & \raisebox{-0.1cm}{0} & \raisebox{-0.1cm}{1} & \raisebox{-0.1cm}{1} & \raisebox{-0.1cm}{3} & \raisebox{-0.1cm}{4} & \raisebox{-0.1cm}{9} & \raisebox{-0.1cm}{14} & \raisebox{-0.1cm}{27} & \raisebox{-0.1cm}{44} & \raisebox{-0.1cm}{80} & \raisebox{-0.1cm}{132} & \raisebox{-0.1cm}{232} & \raisebox{-0.1cm}{?} \\ [1.5ex]
    \hline
    \raisebox{-0.1cm}{$\dim (\mathbb{G}_d^{\mathfrak{sl}_N})$} & \raisebox{-0.1cm}{0} & \raisebox{-0.1cm}{1} & \raisebox{-0.1cm}{1} & \raisebox{-0.1cm}{3} & \raisebox{-0.1cm}{4} & \raisebox{-0.1cm}{\textbf{8}} & \raisebox{-0.1cm}{\textbf{11}} & \raisebox{-0.1cm}{\textbf{19}} & \raisebox{-0.1cm}{\textbf{25}} & \raisebox{-0.1cm}{\textbf{39}} & \raisebox{-0.1cm}{\textbf{50}} & \raisebox{-0.1cm}{\textbf{75}} & \raisebox{-0.1cm}{\textbf{95}} \\ [1.5ex]
    \hline
\end{tabular}
\end{center}\caption{\footnotesize Numbers of Vassiliev invariants and the HOMFLY group factors. The numbers of Vassiliev invariants at orders $\geq 13$ are still unknown. Starting from the 6-th order, the number of $\mathfrak{sl}_N$ group factors becomes less than the number of Vassiliev invariants.
} 
\end{table}

In some models of quantum field theory, correlators satisfy certain relations or equations that can be solved, allowing one to find the correlators without calculating the path integral. Sometimes this can only be done in a specific sector of the theory, which is usually due to the presence of additional (hidden) symmetries in this sector. Examples include spin chains~\cite{sato1978holonomic}, two-dimensional gravity~\cite{gross1990nonperturbative}, Seiberg-Witten theory~\cite{seiberg1994monopoles,seiberg1994electric}, two-dimensional conformal field theories, and matrix models. Sometimes these methods not only help to calculate correlators but also yield some nonperturbative results. Since Chern--Simons theory is topological, one might expect that it also has similar relations on correlators. Some of these are known. For example, Garoufalidis~\cite{garoufalidis2005colored} proved that $H_R^{\cal L}$ satisfies linear difference equations for symmetric representations $R$. However, a generalization to asymmetric representations and other Lie algebras is not yet known. Another example is related to torus knots. In this case, for $H_R^{\cal L}$ (for arbitrary representations), there exists a matrix model whose correlators are related by loop equations. Moreover, they form a so-called topological recursion; see~\cite{brini2012torus,dunin2019combinatorial,dunin2022topological} for details. Many interesting properties emerge there, but this approach has not yet been generalized to non-torus knots. In our paper, we show that certain non-trivial algebraic relations can always be written. We do not prove this in some generic cases, but we discuss numerous examples.

A full set of observables in the Chern--Simons theory is formed by its correlators being Vassiliev invariants. In this paper, we study which correlators in Chern--Simons theory are independent, what reduces to the question of independent Vassiliev invariants. More precisely, Vassiliev invariants form algebra with respect to ordinary multiplication of functions. Any element of this algebra can be uniquely represented as a polynomial in basis of primary elements \cite{chmutov2012introduction}. However, there are two types of relations in this algebra, 1{\rm T} and 4{\rm T} relations, which greatly complicate the search for primary elements. Today, they are known only up to, and including, order 13 \cite{lanina2022implications}.

However, in practice, we do not always need to work in such generality, since we usually work with some subset of knots in which the knots are somehow parameterized. Usually, one considers planar diagrams of knots in which the parameters are numbers of crossings. We call such subsets of knots \textit{families}. Well-known families of knots are \textit{torus} knots, \textit{twist} knots (Fig.\,\ref{fig:tw-knot}) and \textit{pretzels}, see Fig.\,\ref{fig:pretzel}.

For any such knot family, additional relations on Vassiliev invariants appear, which often depend on the topology of a given family. However, they still have a number of common properties. In this paper, we study what new relations on the Vassiliev invariants, and as consequence on the correlators of the Chern--Simons theory, appear in various knot families.


\bigskip

The paper is organized as follows. In Section~\ref{sec:preliminaries}, we introduce objects and their properties that we use in our study. The mathematical definition of Vassiliev invariants based on the so-called Vassiliev skein relation (see Fig.\,\ref{fig:Vass-skein}) is given in Section~\ref{sec:VI}. In Section~\ref{sec:VI-prop}, we provide topological properties of Vassiliev invariants that we use to fix Vassiliev invariants for some knot families. We show that the Vassiliev skein relation allows one to calculate Vassiliev invariants of any order for some knot families in Section~\ref{sec:VI-ex}. In Section~\ref{sec:considered-knots}, we present knot families with Vassiliev invariants polynomial in knot parameters, which we consider in our study. We state main results of this paper in Section~\ref{sec:main-results}. First, the number of algebraically independent polynomial Vassiliev invariants of any $k$-parametric family is $\leq k$. Second, we have conjectured that any 1-parametric algebra of polynomial Vassiliev invariants is finitely generated while in the case of more parameters, we have provided example of infinitely generated algebra. The third claim is that complete knot invariant for some knot families consists of $k$ Vassiliev invariants. In subsequent sections, we explain these results in detail and examples.   






\setcounter{equation}{0}
\section{Notations and basic terminology}\label{sec:preliminaries}

In this section, we provide basic definitions and properties of objects that we use in our study.

\subsection{Vassiliev invariants}\label{sec:VI} 

In our analysis, we use mathematical definition of Vassiliev knot invariants, so in this section, we introduce needed definitions, see for example~\cite{chmutov2012introduction}. 

\begin{definition}
    A map $f$ of a one-dimensional manifold to $\mathbb{R}^3$ is called curve.
\end{definition}

\begin{definition}
    A point $p$ is called double point of a curve $f$ if in a neighbourhood of the point $p$ the curve $f$ has two branches with non-collinear tangents, see Fig.\ref{fig:double-point}.
\end{definition}

\begin{figure}[h!]
    \begin{picture}(100,50)(-225,-25)

\thicklines

\qbezier(-15,15)(20,7)(30,-20)
\qbezier(-10,-10)(20,7)(50,0)
\put(16,0){{\color{black}\circle*{5}}}

\thinlines

\put(16,0.5){\vector(1,0.2){30}}
\put(16,1.25){\vector(-1,0.75){23}}
    
\end{picture}
    \caption{\footnotesize A double point.}
    \label{fig:double-point}
\end{figure}

\begin{definition}
    A knot is called singular if it has at least one double point.
\end{definition}

\noindent 
Any knot invariant can be extended to singular knots by means of the Vassiliev skein relation depicted in Fig.\,\ref{fig:Vass-skein}.

\begin{figure}[h!]
    \begin{picture}(300,30)(-170,0)

\put(-29,10){\mbox{$ {\cal V} \ \Bigg( $}}
\put(15,10){\mbox{$\Bigg) $}}
\put(-11,0){\vector(1,1){24}}
\put(13,0){\line(-1,1){14}}
\put(-1,14){\vector(-1,1){10}}
\put(1,12){{\color{black}\circle*{4}}}

\put(66,0){
\put(-35,10){\mbox{$= \ \ {\cal V} \ \Bigg($}}
\put(29,10){\mbox{$\Bigg)$}}

\put(1,0){\vector(1,1){24}}
\put(25,0){\line(-1,1){10}}
\put(11,14){\vector(-1,1){10}}

}

\put(148,0){
\put(-39,10){\mbox{$- \ \ {\cal V} \ \Bigg($}}
\put(23,10){\mbox{$\Bigg)$}}

\put(19,0){\vector(-1,1){24}}
\put(-5,0){\line(1,1){10}}
\put(9,14){\vector(1,1){10}}
}

\end{picture}
    \caption{\footnotesize Vassiliev skein relation.}
    \label{fig:Vass-skein}
\end{figure}

\begin{definition}\label{def:VI}
    A knot invariant is said to be a Vassiliev invariant of order $\leq n$ if its extension vanishes on all singular knots with $\geq n+1$ double points. A Vassiliev invariant is said to be of order $n$ if it is of order $\leq n$ but not of order $\leq n-1$.
\end{definition}

\noindent
It was proven in~\cite{gusarov1991new,birman1993knot,bar1995vassiliev,birman1976stable} that correlators in the Chern--Simons theory~\eqref{V-inv} are Vassiliev invariants by this definition. 
In general, Vassiliev invariants may take values in an arbitrary abelian group. In this paper, we only deal with the Vassiliev invariants coming from the Chern--Simons correlation functions~\eqref{V-inv} which are just rational numbers.

The set of all Vassiliev invariants forms commutative filtered graded algebra with respect to the usual multiplication of functions. Crucial role in the algebra of Vassiliev invariants is played by its {\it primitive} elements ${\cal V}_{\rm prim}$ which are additive with respect to the knots composition operation:
\begin{equation}\label{primitive}
    {\cal V}_{\rm prim}^{{\cal K}_1\#{\cal K}_2} = {\cal V}_{\rm prim}^{{\cal K}_1} + {\cal V}_{\rm prim}^{{\cal K}_2}\,.
\end{equation}
Primitive Vassiliev invariants are generators in the algebra of Vassiliev invariants. If now, we descend from the set of all knots to a knot family, additional relations on primitive Vassiliev invariants can appear, and their subset will generate the algebra for this fixed knot family. In this case, we call generators of such descended algebra as {\it primary} Vassiliev invariants, so that any Vassiliev invariant is a polynomial in primary Vassiliev invariants. Note that sometimes there are still algebraic relations between these generators (see for example~\eqref{alg-rel-2-str-torus}), and sometimes there are not, depending on the knot family under consideration.

In other words, for the whole algebra of Vassiliev invariants, generators = primary elements = primitive elements, while inside a knot family, this notion splits. All primitive elements still obey~\eqref{primitive}, but the algebra is generated by a smaller subset of primary Vassiliev invariants.








\subsection{Properties of Vassiliev invariants}\label{sec:VI-prop}

In this section, we state several important properties of Vassiliev invariants which we use to fix coefficients in polynomial Vassiliev invariants.

\paragraph{Free additive and multiplicative constants.} The 
fact that Vassiliev skein relation (see Fig.\,\ref{fig:Vass-skein}) is difference implies that Vassiliev invariants are defined up to an additive constant. 
It is also clear that they also have unfixed multiplicative constant.

\paragraph{Fixing additive constants.} In the 3d Chern--Simons theory, additive free constant can be fixed for knot families including the unknot by imposing the following normalization:
\begin{equation}
    {\cal H}^{\bigcircle}_R = 1\,.
\end{equation}
Then, the perturbative expansion~\eqref{LoopExpansionHOMFLY} implies that
\begin{equation}\label{Vass-unknot}
    {\cal V}_0^{\bigcircle}=1\,, \quad {\cal V}^{\bigcircle}_{d\,\geq\, 0}=0\,,
\end{equation}
where ${\cal V}_{d}^{\bigcircle}$ is Vassiliev invariant for the unknot of order $\leq d$.

Note that sometimes the multiplicative constant is fixed by the value on the trefoil knot $3_1$~\cite{willerton2002first,dunin2013kontsevich}. Namely, the second and the third Vassiliev invariants multiplicative constant is fixed by demanding ${\cal V}_2^{3_1}=1$, ${\cal V}_3^{3_1}=1$.

\paragraph{Symmetry under mirror image.} Wilson loops in the 3-dimensional Chern--Simons theory possess the following property for knot $\overline{\cal K}$ mirror to a knot $\cal K$:
\begin{equation}
    {\cal H}^{\overline{\cal K}}_R(q,A) = {\cal H}_R^{\cal K}(q^{-1},A^{-1})\,,
\end{equation}
which for the perturbative series~\eqref{LoopExpansionHOMFLY}, implies the equality
\begin{equation}
    \sum_{n=0}^{\infty} \hbar^{n} \sum_{m=1}^{\dim\left(\mathbb{G}_{n}\right)} \mathcal{V}^{\overline{\mathcal{K}}}_{n,m}\mathcal{G}_{n, m}^R = \sum_{n=0}^{\infty} (-\hbar)^{n} \sum_{m=1}^{\dim\left(\mathbb{G}_{n}\right)} \mathcal{V}^{\mathcal{K}}_{n,m}\mathcal{G}_{n, m}^R\,.
\end{equation}
And therefore, Vassiliev invariants for mirror knot are
\begin{equation}\label{Vass-mirror}
    \mathcal{V}^{\overline{\mathcal{K}}}_{n,m} = (-1)^n \, \mathcal{V}^{\mathcal{K}}_{n,m}\,.
\end{equation}
A mathematical proof that all Vassiliev invariants obey this property can be found in~\cite{chmutov2012introduction}.

\paragraph{Completeness conjecture.} A whole set of Vassiliev invariants is supposed to form complete knot invariant.

\medskip

\noindent This conjecture was proposed in the original paper \cite{Vassiliev} in 1990, but remains open. It has neither been proven nor disproved despite numerous efforts, including computer searches for counterexamples. However, the existence of such a large number of independent numerical invariants seems unnecessary. Indeed, nonequivalent knots form a discrete, countable set. This means they can be simply numbered, thereby revealing a complete numerical invariant defined by such a numbering. This suggests that Vassiliev invariants should not only distinguish knots but also explicitly describe their various topological properties.

\subsection{Examples of derivation of Vassiliev invariants for simplest knot families}\label{sec:VI-ex}

In Section~\ref{sec:VI}, we have defined Vassiliev invariant to satisfy Vassiliev skein relation from Fig.\,\ref{fig:Vass-skein}. In this section, we show that, using this relation, topology of a knot family and properties of Vassiliev invariants~\eqref{Vass-unknot},~\eqref{Vass-mirror}, one can completely fix Vassiliev invariants for two-strand (Section~\ref{sec:T[2,n]-VI}) and twist knots (Section~\ref{sec:Tw-VI}). We also show that Vassiliev invariants for a generic 2-strand evolutionary family are polynomials in knot parameters, see Section~\ref{sec:gen-2-srt-ev-family}. 


\subsubsection{Two-strand torus knots}\label{sec:T[2,n]-VI}

\begin{figure}[h!]
\begin{picture}(300,100)(-60,-55)

\put(0,0){\mbox{$ 0 \ \ = \ \ {\cal V}_0 \ \Bigg( $}}

\put(35,0){

\qbezier(60,30)(77,0)(60,-30)
\qbezier(30,30)(13,0)(30,-30)

\qbezier(50,30)(55,35)(60,30)
\qbezier(30,30)(35,35)(40,30)

\put(50,20){\vector(-1,1){12}}
\put(40,20){\vector(1,1){12}}
\put(45,25){{\color{black}\circle*{3.5}}}

\put(40,0){\vector(1,1){12}}
\put(50,0){\line(-1,1){3}}
\put(43,7){\vector(-1,1){5}}
\qbezier(40,10)(35,15)(40,20)
\qbezier(50,10)(55,15)(50,20)

\put(39,-10){\mbox{$ \dots $}}

\put(0,-30){

\put(40,0){\vector(1,1){12}}
\put(50,0){\line(-1,1){3}}
\put(43,7){\vector(-1,1){5}}

\qbezier(50,0)(55,-5)(60,0)
\qbezier(30,0)(35,-5)(40,0)

}
}

\put(110,0){\mbox{$\Bigg) \ \ = \ \ {\cal V}_0 \ \Bigg( $}}

\put(175,-55){\mbox{$T[2,n]$}}

\put(145,0){

\qbezier(60,30)(77,0)(60,-30)
\qbezier(30,30)(13,0)(30,-30)

\qbezier(50,30)(55,35)(60,30)
\qbezier(30,30)(35,35)(40,30)

\put(43,27){\vector(-1,1){5}}
\put(50,20){\line(-1,1){3}}
\put(40,20){\vector(1,1){12}}

\put(40,0){\vector(1,1){12}}
\put(50,0){\line(-1,1){3}}
\put(43,7){\vector(-1,1){5}}
\qbezier(40,10)(35,15)(40,20)
\qbezier(50,10)(55,15)(50,20)

\put(39,-10){\mbox{$ \dots $}}

\put(0,-30){

\put(40,0){\vector(1,1){12}}
\put(50,0){\line(-1,1){3}}
\put(43,7){\vector(-1,1){5}}

\qbezier(50,0)(55,-5)(60,0)
\qbezier(30,0)(35,-5)(40,0)

}
}

\put(220,0){\mbox{$\Bigg) \ \ - \ \ {\cal V}_0 \ \Bigg( $}}

\put(275,-55){\mbox{$T[2,n-2]$}}

\put(255,0){

\qbezier(60,30)(77,0)(60,-30)
\qbezier(30,30)(13,0)(30,-30)

\qbezier(50,30)(55,35)(60,30)
\qbezier(30,30)(35,35)(40,30)

\put(50,20){\vector(-1,1){12}}
\put(40,20){\line(1,1){3}}
\put(47,27){\vector(1,1){5}}

\put(40,0){\vector(1,1){12}}
\put(50,0){\line(-1,1){3}}
\put(43,7){\vector(-1,1){5}}
\qbezier(40,10)(35,15)(40,20)
\qbezier(50,10)(55,15)(50,20)

\put(39,-10){\mbox{$ \dots $}}

\put(0,-30){

\put(40,0){\vector(1,1){12}}
\put(50,0){\line(-1,1){3}}
\put(43,7){\vector(-1,1){5}}

\qbezier(50,0)(55,-5)(60,0)
\qbezier(30,0)(35,-5)(40,0)

}
}

\put(330,0){\mbox{$\Bigg)$}}
    
\end{picture}
    \caption{\footnotesize By definition, the zeroth Vassiliev invariant vanishes on all singular knots and, in particular, on knots with one double point. Being applied to torus knots together with Vassiliev skein relation, this leads to the fact that the zeroth Vassiliev invariant is a constant.}
    \label{fig:Vass-skein-torus-0}
\end{figure}

\begin{figure}[h!]
\begin{picture}(300,200)(-40,-175)

\put(0,0){\mbox{$ 0 \ \ = \ \ {\cal V}_1 \ \Bigg( $}}

\put(35,0){

\qbezier(60,30)(77,-10)(60,-50)
\qbezier(30,30)(13,-10)(30,-50)

\qbezier(50,30)(55,35)(60,30)
\qbezier(30,30)(35,35)(40,30)

\put(50,20){\vector(-1,1){12}}
\put(40,20){\vector(1,1){12}}
\put(45,25){{\color{black}\circle*{3.5}}}

\put(40,0){\vector(1,1){12}}
\put(50,0){\vector(-1,1){12}}
\put(45,5){{\color{black}\circle*{3.5}}}
\qbezier(40,10)(35,15)(40,20)
\qbezier(50,10)(55,15)(50,20)

\put(0,-20){

\put(40,0){\vector(1,1){12}}
\put(50,0){\line(-1,1){3}}
\put(43,7){\vector(-1,1){5}}
\qbezier(40,10)(35,15)(40,20)
\qbezier(50,10)(55,15)(50,20)

}

\put(39,-30){\mbox{$ \dots $}}

\put(0,-50){

\put(40,0){\vector(1,1){12}}
\put(50,0){\line(-1,1){3}}
\put(43,7){\vector(-1,1){5}}

\qbezier(50,0)(55,-5)(60,0)
\qbezier(30,0)(35,-5)(40,0)

}
}

\put(110,0){\mbox{$\Bigg) \ \ = \ \ {\cal V}_1 \ \Bigg( $}}

\put(145,0){

\qbezier(60,30)(77,-10)(60,-50)
\qbezier(30,30)(13,-10)(30,-50)

\qbezier(50,30)(55,35)(60,30)
\qbezier(30,30)(35,35)(40,30)

\put(43,27){\vector(-1,1){5}}
\put(50,20){\line(-1,1){3}}
\put(40,20){\vector(1,1){12}}

\put(40,0){\vector(1,1){12}}
\put(50,0){\vector(-1,1){12}}
\put(45,5){{\color{black}\circle*{3.5}}}
\qbezier(40,10)(35,15)(40,20)
\qbezier(50,10)(55,15)(50,20)

\put(0,-20){

\put(40,0){\vector(1,1){12}}
\put(50,0){\line(-1,1){3}}
\put(43,7){\vector(-1,1){5}}
\qbezier(40,10)(35,15)(40,20)
\qbezier(50,10)(55,15)(50,20)

}

\put(39,-30){\mbox{$ \dots $}}

\put(0,-50){

\put(40,0){\vector(1,1){12}}
\put(50,0){\line(-1,1){3}}
\put(43,7){\vector(-1,1){5}}

\qbezier(50,0)(55,-5)(60,0)
\qbezier(30,0)(35,-5)(40,0)

}
}

\put(220,0){\mbox{$\Bigg) \ \ - \ \ {\cal V}_1 \ \Bigg( $}}

\put(255,0){

\qbezier(60,30)(77,-10)(60,-50)
\qbezier(30,30)(13,-10)(30,-50)

\qbezier(50,30)(55,35)(60,30)
\qbezier(30,30)(35,35)(40,30)

\put(50,20){\vector(-1,1){12}}
\put(40,20){\line(1,1){3}}
\put(47,27){\vector(1,1){5}}

\put(40,0){\vector(1,1){12}}
\put(50,0){\vector(-1,1){12}}
\put(45,5){{\color{black}\circle*{3.5}}}
\qbezier(40,10)(35,15)(40,20)
\qbezier(50,10)(55,15)(50,20)

\put(0,-20){

\put(40,0){\vector(1,1){12}}
\put(50,0){\line(-1,1){3}}
\put(43,7){\vector(-1,1){5}}
\qbezier(40,10)(35,15)(40,20)
\qbezier(50,10)(55,15)(50,20)

}

\put(39,-30){\mbox{$ \dots $}}

\put(0,-50){

\put(40,0){\vector(1,1){12}}
\put(50,0){\line(-1,1){3}}
\put(43,7){\vector(-1,1){5}}

\qbezier(50,0)(55,-5)(60,0)
\qbezier(30,0)(35,-5)(40,0)

}
}

\put(330,0){\mbox{$\Bigg)$}}

\put(350,0){\mbox{$=$}}


\put(0,-100){

\put(50,-75){\mbox{$T[2,n]$}}

\put(0,0){\mbox{$ = \ \ {\cal V}_1 \ \Bigg( $}}

\put(20,0){

\qbezier(60,30)(77,-10)(60,-50)
\qbezier(30,30)(13,-10)(30,-50)

\qbezier(50,30)(55,35)(60,30)
\qbezier(30,30)(35,35)(40,30)

\put(43,27){\vector(-1,1){5}}
\put(50,20){\line(-1,1){3}}
\put(40,20){\vector(1,1){12}}

\put(40,0){\vector(1,1){12}}
\put(50,0){\line(-1,1){3}}
\put(43,7){\vector(-1,1){5}}
\qbezier(40,10)(35,15)(40,20)
\qbezier(50,10)(55,15)(50,20)

\put(0,-20){

\put(40,0){\vector(1,1){12}}
\put(50,0){\line(-1,1){3}}
\put(43,7){\vector(-1,1){5}}
\qbezier(40,10)(35,15)(40,20)
\qbezier(50,10)(55,15)(50,20)

}

\put(39,-30){\mbox{$ \dots $}}

\put(0,-50){

\put(40,0){\vector(1,1){12}}
\put(50,0){\line(-1,1){3}}
\put(43,7){\vector(-1,1){5}}

\qbezier(50,0)(55,-5)(60,0)
\qbezier(30,0)(35,-5)(40,0)

}
}

\put(150,-75){\mbox{$T[2,n-2]$}}

\put(95,0){\mbox{$\Bigg) \ \ - \ \ {\cal V}_1 \ \Bigg( $}}

\put(130,0){

\qbezier(60,30)(77,-10)(60,-50)
\qbezier(30,30)(13,-10)(30,-50)

\qbezier(50,30)(55,35)(60,30)
\qbezier(30,30)(35,35)(40,30)

\put(43,27){\vector(-1,1){5}}
\put(50,20){\line(-1,1){3}}
\put(40,20){\vector(1,1){12}}

\put(50,0){\vector(-1,1){12}}
\put(40,0){\line(1,1){3}}
\put(47,7){\vector(1,1){5}}
\qbezier(40,10)(35,15)(40,20)
\qbezier(50,10)(55,15)(50,20)

\put(0,-20){

\put(40,0){\vector(1,1){12}}
\put(50,0){\line(-1,1){3}}
\put(43,7){\vector(-1,1){5}}
\qbezier(40,10)(35,15)(40,20)
\qbezier(50,10)(55,15)(50,20)

}

\put(39,-30){\mbox{$ \dots $}}

\put(0,-50){

\put(40,0){\vector(1,1){12}}
\put(50,0){\line(-1,1){3}}
\put(43,7){\vector(-1,1){5}}

\qbezier(50,0)(55,-5)(60,0)
\qbezier(30,0)(35,-5)(40,0)

}
}

\put(260,-75){\mbox{$T[2,n-2]$}}

\put(205,0){\mbox{$\Bigg) \ \ - \ \ {\cal V}_1 \ \Bigg( $}}

\put(240,0){

\qbezier(60,30)(77,-10)(60,-50)
\qbezier(30,30)(13,-10)(30,-50)

\qbezier(50,30)(55,35)(60,30)
\qbezier(30,30)(35,35)(40,30)

\put(50,20){\vector(-1,1){12}}
\put(40,20){\line(1,1){3}}
\put(47,27){\vector(1,1){5}}

\put(40,0){\vector(1,1){12}}
\put(50,0){\line(-1,1){3}}
\put(43,7){\vector(-1,1){5}}
\qbezier(40,10)(35,15)(40,20)
\qbezier(50,10)(55,15)(50,20)

\put(0,-20){

\put(40,0){\vector(1,1){12}}
\put(50,0){\line(-1,1){3}}
\put(43,7){\vector(-1,1){5}}
\qbezier(40,10)(35,15)(40,20)
\qbezier(50,10)(55,15)(50,20)

}

\put(39,-30){\mbox{$ \dots $}}

\put(0,-50){

\put(40,0){\vector(1,1){12}}
\put(50,0){\line(-1,1){3}}
\put(43,7){\vector(-1,1){5}}

\qbezier(50,0)(55,-5)(60,0)
\qbezier(30,0)(35,-5)(40,0)

}
}

\put(370,-75){\mbox{$T[2,n-4]$}}

\put(315,0){\mbox{$\Bigg) \ \ + \ \ {\cal V}_1 \ \Bigg( $}}

\put(350,0){

\qbezier(60,30)(77,-10)(60,-50)
\qbezier(30,30)(13,-10)(30,-50)

\qbezier(50,30)(55,35)(60,30)
\qbezier(30,30)(35,35)(40,30)

\put(50,20){\vector(-1,1){12}}
\put(40,20){\line(1,1){3}}
\put(47,27){\vector(1,1){5}}

\put(50,0){\vector(-1,1){12}}
\put(40,0){\line(1,1){3}}
\put(47,7){\vector(1,1){5}}
\qbezier(40,10)(35,15)(40,20)
\qbezier(50,10)(55,15)(50,20)

\put(0,-20){

\put(40,0){\vector(1,1){12}}
\put(50,0){\line(-1,1){3}}
\put(43,7){\vector(-1,1){5}}
\qbezier(40,10)(35,15)(40,20)
\qbezier(50,10)(55,15)(50,20)

}

\put(39,-30){\mbox{$ \dots $}}

\put(0,-50){

\put(40,0){\vector(1,1){12}}
\put(50,0){\line(-1,1){3}}
\put(43,7){\vector(-1,1){5}}

\qbezier(50,0)(55,-5)(60,0)
\qbezier(30,0)(35,-5)(40,0)

}
}

\put(425,0){\mbox{$\Bigg)$}}

}
    
\end{picture}
    \caption{\footnotesize In the first row we write the condition that the first Vassiliev invariant turns to zero on a torus knot with two double points. Then, we resolve the upper double point with the use of Vassiliev skein relation in Fig.\,\ref{fig:Vass-skein}. In the second line, we apply Vassiliev skein relation to the remaining double point and obtain the difference relation that restricts the first Vassiliev invariant for a torus knot $T[2,n]$ to be linear in $n$.}
    \label{fig:Vass-skein-torus-1}
\end{figure}


First, consider the family of 2-strand torus knots $T[2,n]$ with odd $n$. In Figs.\,\ref{fig:Vass-skein-torus-0},\,\ref{fig:Vass-skein-torus-1}, we utilize vanishing of Vassiliev invariants on corresponding singular knots and Vassiliev skein relation from Fig.\,\ref{fig:Vass-skein} to obtain relations for the 0-th and 1-st Vassiliev invariants for different 2-strand torus knots. These relations turn out to be the following requirements for difference derivatives:
\begin{equation}
\begin{aligned}
    \frac{d{\cal V}_0^{T[2,n]}}{dn} = 0\,, \quad \frac{d^2{\cal V}_1^{T[2,n]}}{dn^2} = 0
\end{aligned}
\end{equation}
meaning that
\begin{equation}
    {\cal V}_0^{T[2,n]} = a_{0,0}\,,\quad {\cal V}_1^{T[2,n]} = a_{1,0} + a_{1,1} n\,.
\end{equation}
It can be analogously shown that, in general, we have the equation
\begin{equation}
    \frac{d^{d+1}{\cal V}_d^{T[2,n]}}{dn^{d+1}} = 0
\end{equation}
with the solution\footnote{Note that in general we have more than one Vassiliev invariant at an order $d$. When writing expressions for just ${\cal V}_d$, we mean that they hold for all ${\cal V}_{d,i}$, $i=1,\dots,\dim(\mathcal{A}_d)$.}
\begin{equation}\label{v-tor-gen}
    {\cal V}^{T[2,n]}_d = \sum\limits_{j=0}^d a_{d,j}\, n^j\,,
\end{equation}
where now $a_{d,j}$ are unfixed rational numbers. To fix some coefficients $a_{d,j}$, one can use the topological properties of two-strand torus knots:
\begin{equation}\label{tor-top}
    T[2,-n]=\overline{T[2,n]}\,,\quad T[2,1]={\rm unknot}\,,
\end{equation}
where the overlining means taking the mirror image of knot, and the properties of Vassiliev invariants~\eqref{Vass-unknot},~\eqref{Vass-mirror} derived in the previous section. This method allows one to calculate the Vassiliev invariants of arbitrary orders. Note that all Vassiliev invariants are proportional to $n^2-1$ due to the fact that $T[2,1]=T[2,-1]={\rm unknot}$ and~\eqref{Vass-unknot}. Vassiliev invariants of odd order are also proportional to $n$ due to the mirror symmetry ${\cal V}^{T[2,n]}_{2p+1}=-{\cal V}^{T[2,-n]}_{2p+1}$ which follows from~\eqref{tor-top},~\eqref{Vass-mirror}. No other restrictions are imposed, thus, the remaining factor in the 2-strand torus Vassiliev invariants is an arbitrary polynomial in $n^2$ of a proper degree. This fact implies that the number of linearly independent Vassiliev invariants ${\cal V}^{T[2,n]}_{2p}$ and ${\cal V}^{T[2,n]}_{2p+1}$ is equal to $p$ for $p\neq 0$, and allows one to write down all these Vassiliev invariants of a fixed order explicitly:
\begin{equation}\label{T[2,n]-VI}
\begin{aligned}
    {\cal V}^{T[2,n]}_{0} &= 1\,,\quad {\cal V}^{T[2,n]}_{1} = 0\,, \\
    {\cal V}^{T[2,n]}_{2p} &= (n^2-1)\sum_{k=0}^{p-1} \alpha_{2p,2k} \, n^{2k}\,,\quad p\neq 0\,, \\
    {\cal V}^{T[2,n]}_{2p+1} &= n(n^2-1)\sum_{k=0}^{p-1} \alpha_{2p+1,2k} \, n^{2k}\,, \quad p\neq 0\,.
\end{aligned}
\end{equation}
Here $\alpha_{d,i}\in \mathbb{Q}$ are arbitrary coefficients.

\subsubsection{Twist knots}\label{sec:Tw-VI}

\begin{figure}[h!]
	\begin{center}
		\includegraphics[width=5.5cm]{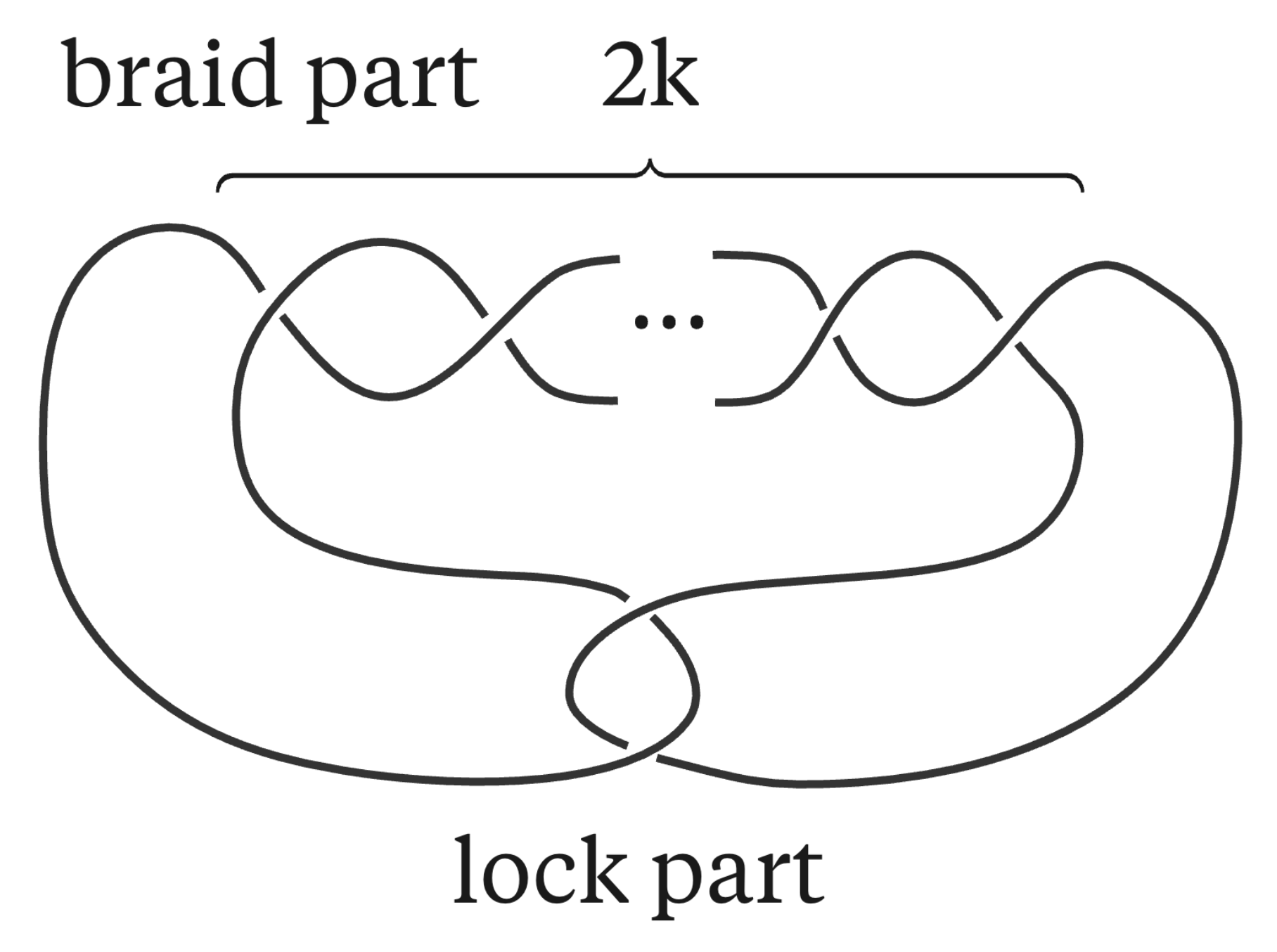}
	\end{center}
    \caption{\footnotesize Twist knot Tw$_{2k}$.}
	\label{fig:tw-knot}
\end{figure}

\noindent  Second, consider the family of twist knots Tw$_{2k}$. Twist knot consists of two parts that we refer to as the braid and lock parts, see Fig.\,\ref{fig:tw-knot}. It is clear that if we put double points only on the braid part of a twist knot, we obtain the same condition of vanishing of the corresponding derivative:
\begin{equation}
    \frac{d^{d+1}{\cal V}_d^{{\rm Tw}_{2k}}}{dk^{d+1}} = 0\,.
\end{equation}
If, now, one puts one double point on the lock part and other ones on the braid part, one obtains the relation of vanishing of the $d$-th Vassiliev invariants on twist knots with $d$ double points on the braid part which, as we have learned, leads to the relation
\begin{equation}
    \frac{d^{d}{\cal V}_d^{{\rm Tw}_{2k}}}{dk^{d}} = 0\,.
\end{equation}
The solution of this equation is
\begin{equation}\label{v-tw-gen}
    {\cal V}^{{\rm Tw}_{2k}}_d = \sum\limits_{j=0}^{d-1} b_{d,j}\, k^j\,.
\end{equation}
One can notice that there exists another relation on Vassiliev invariants for twist knots coming from the case when two double points are put on the lock part. However, the corresponding condition is fulfilled identically for Vassiliev invariants of form~\eqref{v-tw-gen}.

To fix some coefficients $b_{d,j}$, one can use the following connection with the unknot:
\begin{equation}\label{tor-top}
    {\rm Tw}_0={\rm unknot}\,,
\end{equation}
what makes all the Vassiliev invariants to have the factor $k$. Thus, we arrive to the final formula:
\begin{equation}\label{v-tw-gen}
    {\cal V}^{{\rm Tw}_{2k}}_{d>0} = \sum\limits_{j=1}^{d-1} b_{d,j}\, k^j\,,
\end{equation}
so that the Vassiliev invariants of order $\leq d$ consist of $d-1$ linearly independent summands of type $k^l$, $l=1,\dots,d-1$.

\subsubsection{Generic 2-strand evolutionary family}\label{sec:gen-2-srt-ev-family}

Now, consider an arbitrary knot $\cal K$ with $k$ crossings. Instead of each single crossing, insert the 2-braid with $n_i$, $i=1,\dots,k$, crossings and denote this knot as ${\cal K}_{n_1,\dots,\,n_k}$. We call such resulting family as 2-strand evolutionary family. Putting $d+1$ double points on the same 2-braid with $n_i$ crossings, one obtains the following relation:
\begin{equation}
    \frac{\partial^{d+1}{\cal V}_d^{{\cal K}_{n_1,\dots,\,n_k}}}{\partial n_i^{d+1}} = 0\,.
\end{equation}
Obviously, this relation holds for all $n_i$. Thus, the Vassiliev invariants have the form
\begin{equation}\label{pol-VI}
    {\cal V}_d^{{\cal K}_{n_1,\dots,\,n_k}} = \sum\limits_{j_1,\dots,\,j_k=0}^d a_{d,\,j_1,\dots,\,j_k}\,\prod\limits_{i=1}^k n_i^{j_i}\,. 
\end{equation}
There can be further restrictions on these Vassiliev invariants coming from topological properties of a concrete 2-strand evolutionary family. Important for us quality is the polynomiality in knot evolution parameters $n_1,\,n_2,\dots,\,n_k$ of Vassiliev invariants of such families.

\subsection{Knot families under consideration and their Vassiliev invariants}\label{sec:considered-knots}

Our goal is to discover algebraic structures of Vassiliev invariants, i.e. correlators in the 3d Chern--Simons theory. Thus, it is convenient to consider knot families with Vassiliev invariants of type~\eqref{pol-VI} which are polynomials in knot parameters. We have already found out in Section~\ref{sec:gen-2-srt-ev-family} that among such knot families, there are 2-strand evolutionary families. 

\subsubsection{Braiding sequences}

\begin{figure}[h!]
	\begin{center}
		\includegraphics[width=2.8cm]{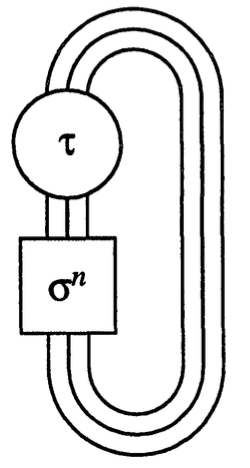}
	\end{center}
    \caption{\footnotesize 1-fold braiding sequence~\cite{stoimenow2000gauss}.}
	\label{fig:br-seq}
\end{figure}

In fact, the variety of families having polynomial Vassiliev invariants is bigger~\cite{stoimenow2000gauss,stoimenow2003vassiliev} and includes knots shown in Fig.\,\ref{fig:br-seq} forming the so-called one parameter {\it braiding sequence}. In this picture, $\tau$ is an arbitrary tangle and $\sigma$ is a pure braid\footnote{Recall that in a pure braid, the beginning and the end of each strand are in the same position.}. If inside a braid, $\sigma_1^{n_1},\dots,\,\sigma_k^{n_k}$ braids with $\sigma_1,\dots,\,\sigma_k$ being pure braids are inserted, the resulting knot family is called $k$-fold braiding sequence, and its Vassiliev invariants are given by~\eqref{pol-VI}.

\subsubsection{Families with same Vassiliev invariants}\label{sec:coinc-VI}

Interesting peculiarity of correlators in the 3-dimensional Chern--Simons theory is that some of them can coincide for the whole knot family. Below we consider some concrete examples of such families.

\begin{figure}[h!]
	\begin{center}
			\includegraphics[width=6.2cm]{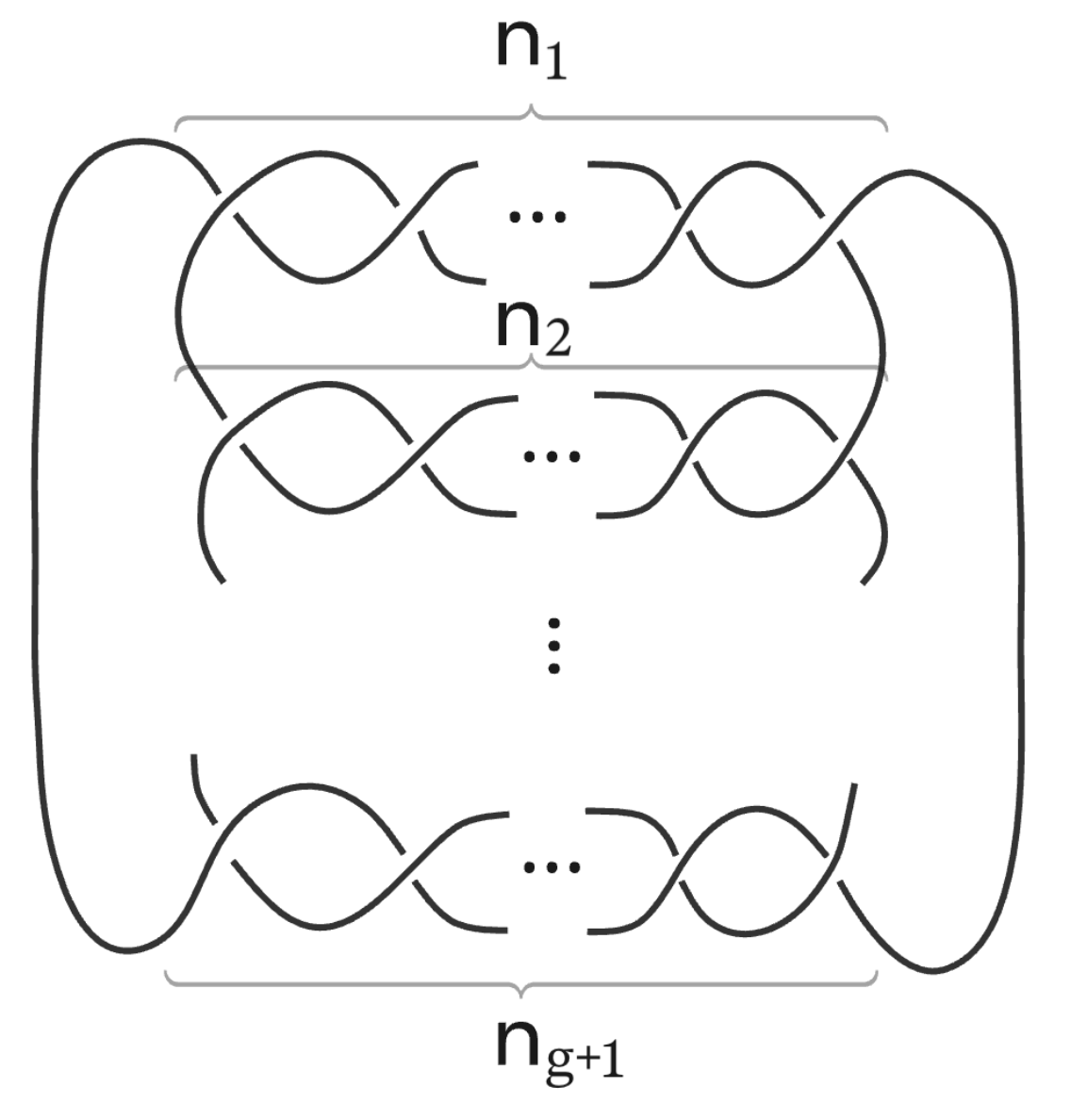}
	\end{center}
    \caption{\footnotesize Pretzel links $P(n_1,\,n_2,\dots,\,n_{g+1})$.}
	\label{fig:pretzel}
\end{figure}

\paragraph{Pretzel links.} The family of pretzel links of genus $g$ is depicted in Fig.\,\ref{fig:pretzel}. There are three possibilities for such links to be one-component, i.e. to be knots:
\begin{itemize}
    \item antiparallel orientation of constituent braids, odd genus $g$, all parameters $n_1,\dots,\,n_{g+1}$ are odd;
    \item parallel orientation of braids, odd genus $g$, one parameter is even and other ones are odd;
    \item one antiparallel braid with even number of crossings, other braids have parallel orientation and odd parameters, even genus $g$.
\end{itemize}

\noindent For all these pretzel knots, the Vassiliev invariants are known up to the 6-th order~\cite{mironov2015colored,sleptsov2016vassiliev}. This fact allows us to find out that the 2-parametric families of parallel pretzel knots $P(n,\,m,\,-n-m,\,1)$ have the vanishing third Vassiliev invariant. Moreover, the subfamily $P(n,\,1,\,-n-1,\,1)$ is amphichiral\footnote{An amphichiral knot is equivalent to its mirror image.} (what can be easily seen by an ambient isotopy), and thus, has all ${\cal V}_{2k+1,i}^{P(n,\,1,\,-n-1,\,1)}\equiv 0$. Also, the antiparallel pretzel knots $P(n,\,m,\,-n,\,-m)$ are amphichiral too.

\begin{figure}[h!]
	\begin{center}
		\includegraphics[width=4.5cm]{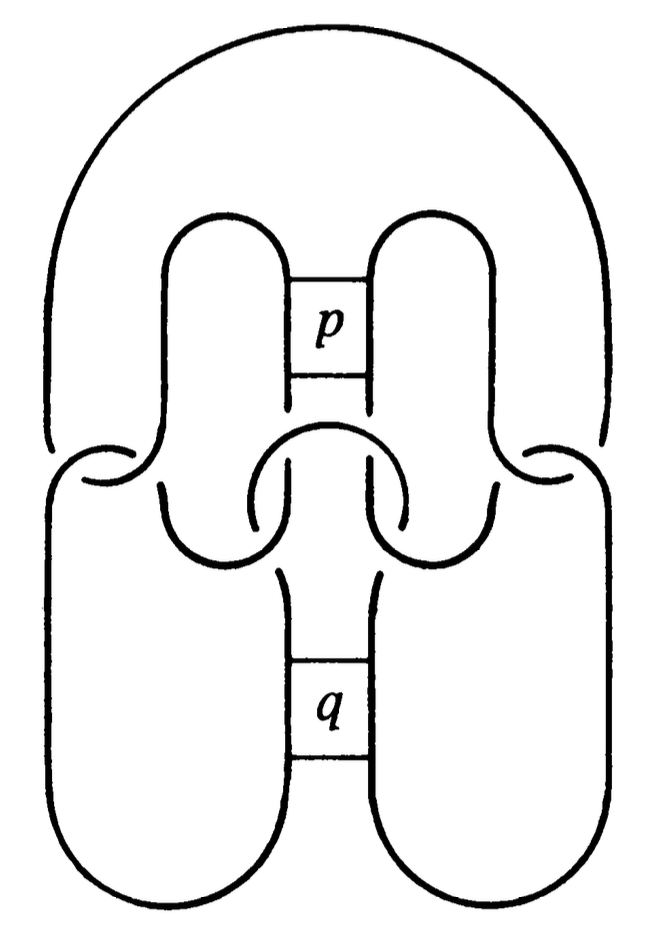}
	\end{center}
    \caption{\footnotesize Kanenobu knots $K(p,q)$~\cite{kanenobu1986infinitely}. A rectangle labeled $n$ stands for $|n|$ full-twists. For fixed $p+q$, the HOMFLY polynomials in the fundamental representation coincide.}
	\label{fig:Kan-knots}
\end{figure}

\paragraph{Kanenobu knots.} The celebrated Kanenobu knots $K(p,q)$~\cite{kanenobu1986infinitely,kanenobu1986examples}, see Fig.\,\ref{fig:Kan-knots}, form knot family having the same fundamental HOMFLY polynomial for a fixed sum of parameters $p+q$. The HOMFLY polynomial in fundamental representation fixes values of the Vassiliev invariants of orders $\leq 4$, thus, at least ${\cal V}_{2,1}^{K(p,q)}$, ${\cal V}_{3,1}^{K(p,q)}$, ${\cal V}_{4,1}^{K(p,q)}$, ${\cal V}_{4,2}^{K(p,q)}$, ${\cal V}_{4,3}^{K(p,q)}$ depend only on the sum of parameters $p+q$. However, one can calculate quantum polynomials in the representation $R=[2]$ and see that all the Vassiliev invariants ${\cal V}_{5,i}^{K(p,q)}$ depend only on $p+q$ too, but the next order Vassiliev invariants already depend both on $p$ and $q$.

\paragraph{Stanford knots.} T. Stanford proved the following theorem~\cite{stanford1996braid,stanford1998vassiliev}. Two knots have the same all Vassiliev invariants of order less than $n$ if and only if they are equivalent modulo the $n$-th group of the lower central series of some pure braid group\footnote{The $n$-th group of the lower central series of the pure braid group $P_k$ is a group ${\rm LCS}_n(P_k)=[{\rm LCS}_{n-1}(P_k),P_k]$ with ${\rm LCS}_1(P_k)=P_k$.}. In Sections~\ref{sec:1-par-v2=0} and~\ref{sec:Stanford}, we use this theorem to examine knots of several first vanishing Vassiliev invariants.

\setcounter{equation}{0}
\section{Main results}\label{sec:main-results}

In this section, we formulate the main results of our paper and the main tools to get these results.

\begin{statement}
    The number of algebraically independent polynomial Vassiliev invariants of any $k$-parametric family is $\leq k$.
\end{statement}

\noindent This statement is consequence of simple fact that more than $k$ polynomials in $k$ variables are algebraically dependent~\cite{lefschetz2005algebraic}.

In fact, we have not found a $k$-parametric knot family with strictly less than $k$ algebraically independent Vassiliev invariants, but nothing seems to contradict this.

\begin{statement}
    There exist $k$-parametric knot families, a complete knot invariant of which consists of $k$ Vassiliev invariants.
\end{statement}

\noindent Indeed, the number of Vassiliev invariants of order $\leq d$ grows with $d$ faster then the number of monomials in these Vassiliev invariants. Thus, at some order, we are able to express all the monomials through Vassiliev invariants. If for a knot family, it is possible to express the family parameters $n_1,\dots,\,n_k$ through Vassiliev invariants in such way, then the corresponding Vassiliev invariants form complete knot family invariant. That is, these Vassiliev invariants distinguish between different knots inside the chosen family.

\begin{statement}
	Algebra of polynomial Vassiliev invariants for a 1-parametric knot family can be finitely generated. In the case of more than one parameter, there exist infinitely generated algebras of Vassiliev invariants. 
\end{statement}

\noindent In the case of a finitely generated algebra, there exists only a finite amount of primary Vassiliev invariants; other ones are non-primary and are polynomials of a proper degree in primary Vassiliev invariants. 


\setcounter{equation}{0}
\section{One-parametric families}\label{sec:1-param}

In this section, we explore relations between Vassiliev invariants that arise in the case of 1-parametric families. In any 1-parametric knot family, we have exactly one algebraically independent Vassiliev invariant. However, there can be more than one primary Vassiliev invariants, but still, their number is finite.

\subsection{Generic one-parametric family}

We explore Vassiliev invariants of an arbitrary order $\leq d$ of a generic one-parametric family:
\begin{equation}\label{v-gen}
    {\cal V}^{{\cal K}_n}_{d,i} = \sum\limits_{j=0}^d a_{d,j}^{(i)}\, n^j\,,\quad i = 1,\dots,\,\dim\left(\mathcal{A}_{d}\right)
\end{equation}
where $a_{d,j}^{(i)}\in \mathbb{Q}$ are some rational numbers. We suppose that in a generic 1-parametric family $a_{d,d}\neq 0$ and fix the normalizations by setting $a_{d,d}=1$. 



\paragraph{Primality.} Let us show that in the case of a generic one-parametric family, there are only three primary Vassiliev invariants because:
\begin{equation}\label{prim-gen-1-par}
\begin{aligned}
    {\cal V}^{{\cal K}_n}_{2,1} &=  n^2+ a_{2,1}^{(1)}\, n+a_{2,0}^{(1)}\,,\\
    {\cal V}^{{\cal K}_n}_{3,1} &=  n^3+a_{3,2}^{(1)}\, n^2+ a_{3,1}^{(1)}\, n+a_{3,0}^{(1)} \,, \\
    \tilde{{\cal V}}^{{\cal K}_n}_{4,2} &:= {\cal V}^{{\cal K}_n}_{4,2} -\left({\cal V}^{{\cal K}_n}_{2,1}\right)^2-\alpha_1 {\cal V}^{{\cal K}_n}_{3,1}-\alpha_2 {\cal V}^{{\cal K}_n}_{2,1}-\alpha_3 = \alpha\, n
\end{aligned}
\end{equation}
with $\alpha$-coefficients expressed through $a$-coefficients: 
\begin{equation}
\begin{aligned}
    \alpha_1&=a_{4,3}^{(2)} -2 a_{2,1}^{(1)} \,, \\
    \alpha_2&=-\left(a_{2,1}^{(1)}\right)^2+2\, a_{3,2}^{(1)} a_{2,1}^{(1)} -2\, a_{2,0}^{(1)} +a_{4,2}^{(2)}-a_{3,2}^{(1)} a_{4,3}^{(2)}\,, \\
    \alpha_3&=\left(a_{2,0}^{(1)}\right)^2+\left(a_{2,1}^{(1)}\right)^2 a_{2,0}^{(1)} -2 a_{2,1}^{(1)} a_{3,2}^{(1)} a_{2,0}^{(1)} - a_{4,2}^{(2)} a_{2,0}^{(1)}+a_{3,2}^{(1)} a_{4,3}^{(2)} a_{2,0}^{(1)} + 2 a_{2,1}^{(1)} a_{3,0}^{(1)} + a_{4,0}^{(2)} - a_{3,0}^{(1)} a_{4,3}^{(2)} \,, \\
    \alpha&= \left(a_{2,1}^{(1)}\right)^3-2 a_{3,2}^{(1)} \left(a_{2,1}^{(1)}\right)^2+2 a_{3,1}^{(1)} a_{2,1}^{(1)} - a_{4,2}^{(2)} a_{2,1}^{(1)} + a_{3,2}^{(1)} a_{4,3}^{(2)} a_{2,1}^{(1)} + a_{4,1}^{(2)} - a_{3,1}^{(1)} a_{4,3}^{(2)} \,.
\end{aligned}    
\end{equation}
It is obvious that all Vassiliev invariants of higher orders are expressed through ${\cal V}^{{\cal K}_n}_{2,1}$, ${\cal V}^{{\cal K}_n}_{3,1}$, ${\cal V}^{{\cal K}_n}_{4,2}$. 

\paragraph{Degenerate cases.} One can also conjecture that the degenerate cases\footnote{We call knot families to be degenerate if all Vassiliev invariants at some orders vanish.} of several vanishing Vassiliev invariants give just a finite number of primary Vassiliev invariants. The logic is as follows. If some Vassiliev invariants, say, ${\cal V}^{{\cal K}_n}_{i_1},\dots,{\cal V}^{{\cal K}_n}_{i_k}$ are zero, then the number of elements in algebra of Vassiliev invariants of order $\leq d$ grows faster than the number of partitions of $d$ that do not contain $i_1,\dots,i_k$ as a part. This number grows faster than the linear growth of $n$-powers inside a Vassiliev invariant of order $\leq d$. Thus, we expect that at some high enough order we are able to express all monomials in $n$ through linear combinations of Vassiliev invariants.

Consider a particular case of ${\cal V}^{{\cal K}_n}_{2,1}\equiv 0$. Then, in general, the number of primary invariants is equal to $5$:
\begin{equation}\label{v2=0}
	\begin{aligned}
		{\cal V}^{{\cal K}_n}_{2,1} &\equiv 0\,, \\
		{\cal V}^{{\cal K}_n}_{3,1} &=  n^3+a_{3,2}^{(1)} n^2+ a_{3,1}^{(1)} n+a_{3,0}^{(1)} \,, \\
		\tilde{{\cal V}}^{{\cal K}_n}_{4,2} &:= {\cal V}^{{\cal K}_n}_{4,2} - a_{4,3}^{(2)}\,{\cal V}^{{\cal K}_n}_{3,1}-a_{4,0}^{(2)}=n^4+\tilde{a}^{(2)}_{4,2}n^2+\tilde{a}^{(2)}_{4,1}n\,, \\
		\tilde{{\cal V}}^{{\cal K}_n}_{4,3} &:= {\cal V}^{{\cal K}_n}_{4,3} - a_{4,3}^{(3)}\,{\cal V}^{{\cal K}_n}_{3,1}-a_{4,0}^{(3)}-\tilde{{\cal V}}^{{\cal K}_n}_{4,2}=\tilde{a}^{(3)}_{4,2}n^2+\tilde{a}^{(3)}_{4,1}n\,, \\
		\tilde{{\cal V}}^{{\cal K}_n}_{5,2} &:={\cal V}^{{\cal K}_n}_{5,2} - a_{5,4}^{(2)}\,\tilde{{\cal V}}^{{\cal K}_n}_{4,2} - a_{5,3}^{(2)}{\cal V}^{{\cal K}_n}_3-\alpha_{5,4}^{(2)}\,\tilde{{\cal V}}^{{\cal K}_n}_{4,3}-a_{5,0}^{(2)}=n^5+\tilde{a}_{5,1}^{(2)} n\,, \\
		\tilde{{\cal V}}^{{\cal K}_n}_{5,3} &:={\cal V}^{{\cal K}_n}_{5,3} - a_{5,4}^{(3)}\,\tilde{{\cal V}}^{{\cal K}_n}_{4,2} - a_{5,3}^{(3)}{\cal V}^{{\cal K}_n}_3-\alpha_{5,4}^{(3)}\,\tilde{{\cal V}}^{{\cal K}_n}_{4,3}-a_{5,0}^{(2)}-\tilde{{\cal V}}^{{\cal K}_n}_{5,2}=\tilde{a}_{5,1}^{(3)} n
	\end{aligned}
\end{equation}
where $\alpha_{k,m}^{(i)}$ and $\tilde{a}_{k,m}^{(i)}$ are expressed via $a_{k,m}^{(i)}$. In the generic case of first $k$ vanishing Vassiliev invariants: ${\cal V}^{{\cal K}_n}_{j,i}\equiv 0$ for all $i$ and $j=2,\dots,\,k$, the number of primary Vassiliev invariants (without ${\cal V}_0$) equals $2k+1$.

\bigskip

\noindent One can wonder if there can appear algebraic relations on Vassiliev invariants that restrict the number of Vassiliev invariants of order $\leq d$ to be $\leq d+1$ for some $d$. Or maybe, we can choose specific Vassiliev invariants which are rarely non-zero and do not contain low powers in $n$. Can this force such a 1-parametric algebra of Vassiliev invariants to be infinitely generated? Examples and reasonings provide the intuition that a 1-parametric algebra of polynomial Vassiliev invariants always has only a finite number of generators. Still, we currently lack strict mathematical proof. 

\paragraph{Algebraic independence.} As we have stated at the beginning of Section~\ref{sec:1-param}, there is only one algebraically independent Vassiliev invariant. All generic one-parametric Vassiliev invariants~\eqref{v-gen} are algebraically expressed through ${\cal V}^{{\cal K}_n}_2$. For example, the relation between ${\cal V}^{{\cal K}_n}_2$ and ${\cal V}^{{\cal K}_n}_3$ is
\begin{equation}
    \beta_{6,4} \left({\cal V}^{{\cal K}_n}_2\right)^2+\beta_{6,2} {\cal V}^{{\cal K}_n}_2+\beta_{6,5} {\cal V}^{{\cal K}_n}_3 {\cal V}^{{\cal K}_n}_2 - \left({\cal V}^{{\cal K}_n}_3\right)^2+\beta_{6,3} {\cal V}^{{\cal K}_n}_3+\left({\cal V}^{{\cal K}_n}_2\right)^3+\beta_{6,0}=0   
\end{equation}
with
\begin{equation}\label{beta-coef}
\begin{aligned}
    \beta_{6,2}&=3 a_{2,0}^2+2 a_{3,2}^2 a_{2,0}-4 a_{3,1} a_{2,0}-2 a_{2,1} a_{3,2} a_{2,0}+a_{3,1}^2+3 a_{2,1} a_{3,0}+a_{2,1}^2 a_{3,1}-2 a_{3,0} a_{3,2}-a_{2,1} a_{3,1} a_{3,2}\,, \\
    \beta_{6,3}&=-a_{2,1}^3+a_{3,2} a_{2,1}^2+3 a_{2,0} a_{2,1}-a_{3,1} a_{2,1}+2 a_{3,0}-2 a_{2,0} a_{3,2}\,, \\
    \beta_{6,4}&=-a_{3,2}^2+a_{2,1} a_{3,2}-3 a_{2,0}+2 a_{3,1}\,, \\
    \beta_{6,5}&=2 a_{3,2}-3 a_{2,1}\,, \\
    \beta_{6,0}&=-a_{2,0}^3-a_{3,2}^2 a_{2,0}^2+2 a_{3,1} a_{2,0}^2+a_{2,1} a_{3,2} a_{2,0}^2-a_{3,1}^2 a_{2,0}-3 a_{2,1} a_{3,0} a_{2,0}-a_{2,1}^2 a_{3,1} a_{2,0}+\\ &+2 a_{3,0} a_{3,2} a_{2,0}+a_{2,1} a_{3,1} a_{3,2} a_{2,0}-a_{3,0}^2+a_{2,1}^3 a_{3,0}+a_{2,1} a_{3,0} a_{3,1}-a_{2,1}^2 a_{3,0} a_{3,2}\,.
\end{aligned}
\end{equation}

\paragraph{Complete invariant.} As we have discussed, the number of Vassiliev invariants always grows faster than the number of monomials inside Vassiliev invariant. Thus, at some high enough order, we are able to express all powers of a knot parameter $n$, and, in particular in a generic case, the parameter itself through the Vassiliev invariants. For example, in the non-degenerate case~\eqref{prim-gen-1-par}, the knot parameter is expressed through the Vassiliev invariant of order $\leq 4$, and in the degenerate case~\eqref{v2=0}, the $n$ parameter is expressed through the Vassiliev invariant of order $\leq 5$. Thus, in general, complete invariant of one-parametric knot family consists of one Vassiliev invariant. Nevertheless, the existence of 1-parametric knot family having all coincident Vassiliev invariants is not excluded. If such a family exists, then the  completeness conjecture of Vassiliev invariants from Section \ref{sec:VI-prop} is false.

\subsection{Example of 2-strand torus family}

Using Vassiliev skein relation, we have fully constrained the form of Vassiliev invariants for 2-strand torus knots $T[2,n]$ of any order~\eqref{T[2,n]-VI}. These invariants possess special symmetry, thus, the number of primary Vassiliev invariants becomes even less than in the generic case and equals two. The primary Vassiliev invariants are the second and the third ones:
\begin{equation}\label{V23-T[2,n]}
    {\cal V}^{T[2,n]}_{2} = n^2 - 1\,,\quad {\cal V}^{T[2,n]}_{3} = n(n^2 - 1)\,.
\end{equation}
In this section in what follows, we omit the superscript $T[2,n]$ implying that all the Vassiliev invariants are considered for 2-strand torus knots. These Vassiliev invariants~\eqref{V23-T[2,n]} generate the whole algebra of Vassiliev invariants for the 2-strand torus family. For example
\begin{equation}
\begin{aligned}
    {\cal V}_{4} &=\alpha_{4,2} {\cal V}_{2}^2+(\alpha_{4,0}+\alpha_{4,2})\,{\cal V}_{2}\,,\\
    {\cal V}_{5} &=\alpha_{5,2}\,{\cal V}_{2}{\cal V}_{3}+(\alpha_{5,0}+\alpha_{5,2})\,{\cal V}_{3}\,,\\
    {\cal V}_{6} &=\alpha_{6,4}\,{\cal V}_{3}^2 +(\alpha_{6,4}+\alpha_{6,2})\,{\cal V}_{2}^2+(\alpha_{6,4}+\alpha_{6,2}+\alpha_{6,0})\,{\cal V}_{2}\,,\\
    {\cal V}_{7} &=\alpha_{7,4}{\cal V}_{2}^2{\cal V}_{3}-(\alpha_{7,2}+2\alpha_{7,4}){\cal V}_{2}{\cal V}_{3}+(\alpha_{7,4}+\alpha_{7,2}+\alpha_{7,0}){\cal V}_{3}\,.
\end{aligned}    
\end{equation}
Only one Vassiliev invariant is algebraically independent. For example, there is the following relation between ${\cal V}_{2}$ and ${\cal V}_{3}$:
\begin{equation}\label{alg-rel-2-str-torus}
    {\cal V}_{3}^2={\cal V}_{2}^3+{\cal V}_{2}^2\,.
\end{equation} 
Vassiliev invariant ${\cal V}_3$ is complete knot invariant for the given knot family, i.e. it distinguishes all knots among the family of two-strand torus knots. This follows from the fact that the polynomial $n(n^2-1)$ is increasing monotone function on the domains $(-\infty,-1]$ and $[1,+\infty)$. On the interval $[-1,+1]$ the parameter $n$ takes only 3 integer values $n=-1, \ 0, \ 1$. All these three values correspond to the unknot. 

On the contrary, ${\cal V}_2$ is not complete invariant, because it does not distinguish mirror knots as it remains invariant under the change of the variable $n \rightarrow -n$.

\subsection{Example of twist family}


The Vassiliev invariants for twist knots ${\rm Tw}_{2k}$ have been derived in Section~\ref{sec:Tw-VI}. Note that ${\cal V}^{{\rm Tw}_{2k}}_{2}$ is complete knot invariant of the family of twist knots ${\rm Tw}_{2k}$. In other words, knowing the values of invariants ${\cal V}^{{\rm Tw}_{2k}}_{2}=k$, one can definitely distinguish the knot ${\rm Tw}_{2k}$ of fixed $k$. Moreover, all the Vassiliev invariants of higher orders are obviously expressed through the Vassiliev invariant of the second order, and the dependence is algebraic. In other words, the only primary Vassiliev invariant in this particular case is the second Vassiliev invariant, and it generates all Vassiliev invariants for the whole family of twist knots.  



\subsection{Example of a knot family with a vanishing Vassiliev invariant}\label{sec:1-par-v2=0}

Let us consider a Stanford 1-parametric knot family with the second vanishing Vassiliev invariant. According to the Stanford theorem~\cite{stanford1996braid,stanford1998vassiliev}, such knots are given by the closure of the following braid: ${\cal K}_n = \overline{p_n b}$ for $k$-braids $p_n$, $b\in \mathcal{B}_k$ such that $\bigcirc = \bar{b}$ and $p_n\in{\rm LCS}_3(P_k)$, see more details in Section~\ref{sec:coinc-VI}. For our example, we take 4-braids $b=\sigma_1\sigma_2\sigma_3$ and
\begin{equation}
    p_n = [[\sigma_1\sigma_2^{2n},\,\sigma_3^{2n}\sigma_2^2],\,\sigma_3^{-2n}\sigma_1^{-2n}]
\end{equation}
where $\sigma_1$, $\sigma_2$, $\sigma_3$ are $\mathcal{B}_4$ braid group generators. We obtain the following low-order Vassiliev invariants for the knot family ${\cal K}_n = \overline{p_n b}$:
\begin{equation}
\begin{aligned}
    {\cal V}_{2,1}^{{\cal K}_n} &\equiv 0\,, \\
    {\cal V}_{3,1}^{{\cal K}_n} &= -6n(n^2-1)\,, \\ 
    {\cal V}_{4,1}^{{\cal K}_n} &= \frac{1}{2}\left( {\cal V}_{2,1}^{{\cal K}_n} \right)^2\equiv 0,\quad  {\cal V}_{4,2}^{{\cal K}_n}=16n(n^3+n^2+n-1),\quad {\cal V}_{4,3}^{{\cal K}_n}= 8n(2n^3-n^2+2n+1)\,.
\end{aligned}
\end{equation}
Among these Vassiliev invariants only two are primary because there is the relation
\begin{equation}
    {\cal V}_{4,3}^{{\cal K}_n} = {\cal V}_{4,2}^{{\cal K}_n} + 4 {\cal V}_{3,1}^{{\cal K}_n}\,.
\end{equation}

\setcounter{equation}{0}
\section{Two-parametric families}

In this section, we analyze algebraic structures of the algebra of Vassiliev invariants of several 2-parametric knot families. Unlike the case of 1-parametric families, in Section~\ref{sec:T[m,n]}, we provide an explicit example of an infinitely generated algebra. 

\subsection{Generic two-parametric family}

We explore Vassiliev invariants of an arbitrary order $\leq d$ of a generic two-parametric family:
\begin{equation}\label{v2-gen}
    {\cal V}^{{\cal K}_{n,m}}_{d,l} = \sum\limits_{j=0}^d \sum\limits_{i=0}^d a_{d,i,j}^{(l)}\,m^i n^j\,,\quad l = 1,\dots,\,\dim\left(\mathcal{A}_{d}\right)
\end{equation}
where $a_{d,i,j}^{(l)}\in \mathbb{Q}$ are some rational numbers. 

\paragraph{Primality.} In this case, the question of primality becomes even more complicated than the 1-parametric one, which has not yet been fully studied. Naively, the number of Vassiliev invariants~\eqref{v2-gen} grows faster than the number of $m,\,n$-monomials inside a Vassiliev invariant. So, there should exist families with finitely generated algebra of Vassiliev invariants. However, we have not managed to find such family because, in generic 2-parametric case, topological symmetries are not enough to find Vassiliev invariant of an arbitrary order. And in the particular case of torus knots $T[m,n]$ for which all the Vassiliev invariants are fixed, the Vassiliev invariants algebra is infinitely generated, see Section~\ref{sec:T[m,n]}.

\paragraph{Algebraic independence.} For any knot family, all two-parametric Vassiliev invariants~\eqref{v2-gen} are algebraically expressed through $\leq 2$ Vassiliev invariants. These two Vassiliev invariants are specific for a given knot family. Two Vassiliev invariants ${\cal V}_d$ and ${\cal V}_l$ polynomial in $m$ and $n$ are algebraically independent\footnote{This theorem actually holds for two arbitrary polynomials, not only for Vassiliev invariants.} iff
\begin{equation}\label{VI-2-dep}
    \frac{\partial {\cal V}_d}{\partial m}\cdot \frac{\partial {\cal V}_l}{\partial n} \neq \frac{\partial {\cal V}_d}{\partial n}\cdot \frac{\partial {\cal V}_l}{\partial m}\,.
\end{equation}
In fact, we have not managed to find 2-parametric knot families having less than 2 algebraically independent Vassiliev invariants. It is a long standing problem to find two knots indistinguishable by the whole set of Vassiliev invariants what would lead to the break down of the Vassiliev invariants completeness conjecture. To find such knot family is even more puzzling task. Thus, we have focused on the problem of finding a 2-parametric knot family with an only one algebraically independent Vassiliev invariant, see Section~\ref{sec:less-AI-VI}. We propose that such a knot family should have some symmetry. This could be an explicit topological symmetry: for example, amphichirality or Stanford symmetry that makes several first Vassiliev invariants match. Hidden symmetries can also manifest. For example, there are Kanenobu knots $K(p,q)$ with the HOMFLY polynomials (in fundamental representation), and thus, the lowest order Vassiliev invariants, dependent only on the sum $p+q$.

\paragraph{Complete invariant.} As we have discussed, in a generic case at some high enough order, we can solve a linear system of equations and express all monomials in a knot family parameters as a linear combination of Vassiliev invariants (being also a Vassiliev invariant). In particular, we encode two knot parameters $m$ and $n$ by two Vassiliev invariants. Thus, a complete invariant of a generic given knot family consists of at most two primary Vassiliev invariants. However, we cannot yet exclude the case when the remaining two Vassiliev invariants turn out to be functionally dependent. But such a case will not lead to any serious consequences.




\subsection{Example of $m$-strand torus family}\label{sec:T[m,n]}

We consider the two-parametric family of torus knots $T[m,n]$. 
A Vassiliev invariant of order $\leq d$ for this family is of the form
\begin{equation}\label{v-T[m,n]}
    {{\cal V}_{d}^{T[m,n]} = \sum_{i=0}^d \sum_{j=0}^d a_{d,i,j} m^i n^j}\,.
\end{equation}
We can further constrain the coefficients $a_{d,i,j}$ with help of the following topological properties:
\begin{equation}
    T[m,n] = T[n,m]\,,\quad T[-m,n] = \overline{T[m,n]}\,,\quad T[m,1] = \text{unknot}\,,
\end{equation}
and the properties of Vassiliev invariants~\eqref{Vass-unknot},~\eqref{Vass-mirror}. 

It follows from $T[m,n] = T[n,m]$ that the coefficients $a_{d,i,j}$ are symmetric under the change of $i,\,j\,$: $a_{d,i,j}=a_{d,j,i}$. The property $T[-m,n] = \overline{T[m,n]}$ means that only coefficients with $n$ and $m$ raised to even powers occur in $\mathcal{V}_{2p}^{T[m,n]}$ decomposition, and only coefficients with $n$ and $m$ raised to odd powers are included in $\mathcal{V}_{2p+1}^{T[m,n]}$. Taking into account the unknot restriction we get that $a_{d,i,j}$ are such that the resulting polynomial is proportional to $(n^2-1)(m^2-1)$. In total, we get that a Vassiliev invariant $\mathcal{V}_{2p}^{T[m,n]}$ is product of $(n^2-1)(m^2-1)$ and an arbitrary symmetric polynomial in $m^2$, $n^2$ of degree $p-1$ in each variable, and a Vassiliev invariant $\mathcal{V}_{2p+1}^{T[m,n]}$ is product of $nm(n^2-1)(m^2-1)$ and an arbitrary symmetric polynomial in $m^2$, $n^2$ of degree $p-1$ in each variable. Thus, the numbers of linear independent Vassiliev invariants in $\mathcal{V}_{2p}^{T[m,n]}$ and $\mathcal{V}_{2p+1}^{T[m,n]}$ are $\frac{1}{2}(p+1)p$. We provide explicit answers for Vassiliev invariants of any order:
\begin{equation}\label{VI-T[m,n]}
\begin{aligned}
    \mathcal{V}_{2p}^{T[m,n]} &= (m^2-1)(n^2-1)\sum_{k=0}^{p-1}\sum_{l=0}^k \alpha_{2p,2k,2l}(m^{2k}n^{2l}+m^{2l}n^{2k}) = \sum_{k=0}^{p-1}\sum_{l=0}^k \alpha_{2p,2k,2l} P_{k,l}(m,n) \,, \\ 
    \mathcal{V}_{2p+1}^{T[m,n]} &= mn(m^2-1)(n^2-1)\sum_{k=0}^{p-1}\sum_{l=0}^k \alpha_{2p+1,2k,2l}(m^{2k}n^{2l}+m^{2l}n^{2k}) = mn\sum_{k=0}^{p-1}\sum_{l=0}^k \alpha_{2p,2k,2l} P_{k,l}(m,n)
\end{aligned}
\end{equation}
where we fixed the free multiplication constants so that $\alpha_{2p,2(p-1),2(p-1)}=1$ and $\alpha_{2p+1,2(p-1),2(p-1)}=1$. We see that $T[m,n]$ symmetries fix Vassiliev invariants completely.

In what follows in this subsection, we again omit the superscript $T[m,n]$ of Vassiliev invariants.

\paragraph{Primality.} One can prove that at each order, we have exactly one primary Vassiliev invariant of the form:
\begin{equation}\label{prim-T[m,n]}
\begin{aligned}
    \text{even order } 2p&:\quad {\cal V}_{2p}^{\rm prim} = (n^2 - 1) (m^2 - 1)(m^{2(p-1)}+n^{2(p-1)})\,, \\
    \text{odd order } 2p+1&:\quad {\cal V}_{2p+1}^{\rm prim} = mn(n^2 - 1) (m^2 - 1)(m^{2(p-1)}+n^{2(p-1)})\,.
\end{aligned}
\end{equation}
Since new primary invariant appears at each order, then the algebra of Vassiliev invariants for torus knots $T[m,n]$ is infinitely generated. 

In order to prove that the Vassiliev invariants~\eqref{prim-T[m,n]} form the full set of primary Vassiliev invariants, we need to show that all the Vassiliev invariants~\eqref{VI-T[m,n]} are expressed only through these primary Vassiliev invariants. Let us demonstrate that $\mathcal{V}_{2p}^{T[m,n]}$ for any $p$ is expressed only through ${\cal V}_{2l}^{\rm prim}$. There is the following relation 
\begin{equation}
    {\cal V}_{2(k+1)}^{\rm prim}{\cal V}_{2(l+1)}^{\rm prim} - \frac{1}{2}{\cal V}_{2}^{\rm prim}{\cal V}_{2(k+l+1)}^{\rm prim} = P_{k+1,l+1}(m,n) + P_{k,l}(m,n) - P_{k+1,l}(m,n) - P_{k,l+1}(m,n)\,,
\end{equation}
so that all $P_{k,l}(m,n)$ are recursively expressed through ${\cal V}_{2l}^{\rm prim}$ and $P_{0,0} = {\cal V}_{2}^{\rm prim}$, $P_{1,0} = P_{0,1} = {\cal V}_{4}^{\rm prim}$. The same calculations (up to the $mn$ multiplier) are relevant for the Vassiliev invariants of odd orders.

Also note that all the Vassiliev invariants~\eqref{prim-T[m,n]} are primary. One can wonder if one can extract a Vassiliev invariant of the form~\eqref{prim-T[m,n]} from their product. For example, a product of such Vassiliev invariants of even orders has the form $(m^2 - 1)(n^2 - 1)(m^{2(p-1)}+n^{2(p-1)}+\dots)$, but one cannot extract the primary Vassiliev invariant ${\cal V}_{2p}^{\rm prim}$ because the term $(m^{2(p-1)}+n^{2(p-1)})$ is connected with other monomials by extra $(m^2 - 1)(n^2 - 1)$ multipliers.

\paragraph{Algebraic independence.} Two Vassiliev invariants ${\cal V}_{2}$ and ${\cal V}_{3}$ are algebraically independent. Algebraic relations for some higher-order Vassiliev invariants are:
\begin{equation}
\begin{aligned}
    {\cal V}_{2}{\cal V}_{4}&-\left(1+\frac{1}{2}\alpha_{4,2,0}\right){\cal V}_{3}^2+\frac{1}{4}\alpha_{4,2,0}{\cal V}_{2}^3-\left(\alpha_{4,0,0}+\frac{1}{2}\alpha_{4,2,0}\right){\cal V}_{2}^2=0\,, \\
    {\cal V}_2^2{\cal V}_{5}&-\left(1+\frac{1}{2}\alpha_{5,2,0}\right){\cal V}_{3}^3+\frac{1}{4}\alpha_{5,2,0}{\cal V}_{3}{\cal V}_{2}^3-(\alpha_{5,0,0}+\frac{1}{2}\alpha_{5,2,0}){\cal V}_{3}{\cal V}_{2}^2=0\,.
\end{aligned}
\end{equation}
\paragraph{Complete invariant.} In contrast with the 2-strand case, neither ${\cal V}_{2}^{T[m,n]}$ nor ${\cal V}_{3}^{T[m,n]}$ is complete knot invariant. The Vassiliev invariant ${\cal V}_2$ does not distinguish mirror knots, and for example, ${\cal V}_{3}^{T[4,7]} = {\cal V}_{3}^{T[2,15]}$. However, it seems that the union of ${\cal V}_{2}^{T[m,n]}$ and ${\cal V}_{3}^{T[m,n]}$ form complete knot invariant. Namely, we have checked up to $m,n=10000$ that the primary Vassiliev invariant ${\cal V}_3({\cal V}_2+1)$ is complete knot invariant.

\subsection{Example of pretzel $P(m,n,\overline{2})$ family}

It is convenient to start with the three-parametric family of mixed pretzel knots $P(\overline{n}_1,n_2,n_3)$ with $\overline{n}_1$ being even and $n_2$, $n_3$ being odd in order to form a knot. For the genus $g=2$, pretzel knots $P(n_1,n_2,n_3)$ are invariant under all permutations of the parameters $n_1,\,n_2,\,n_3\,$. Thus, the Vassiliev invariants are symmetric functions in three variables:
\begin{equation}\label{Vass-pretzel-gen}
    {\cal V}_{d}^{P(n_1,n_2,n_3)}=\sum_{k=0}^d \ \sum_{\lambda:\,l_\lambda \leq 2}a_{d,\lambda}\cdot\chi_\lambda
\end{equation}
where $\chi_\lambda$ are Schur polynomials enumerated by Young diagrams $\lambda$ with no more than 2 rows. One also uses the mirror symmetry:
\begin{equation}\label{pretzel-mirr}
    P(-n_1,-n_2,-n_3)=\overline{P(n_1,n_2,n_3)}
\end{equation}
and the corresponding property of Vassiliev invariants~\eqref{Vass-mirror} to fix some coefficients in~\eqref{Vass-pretzel-gen}. Then, we put $n_1=2$ and utilize the topology of $P(m,n,\overline{2})$:
\begin{equation}
    P(m,n,\overline{2})=T[2,m-2]
\end{equation}
and derived formulas for Vassiliev invariants for 2-strand torus knots~\eqref{T[2,n]-VI}. However, already at the second order, the mentioned symmetries of pretzel knots are not enough to fix Vassiliev invariants entirely, and one should use the known answers for Vassiliev invariants of some concrete pretzel knots~\cite{katlas} or just take formulas from~\cite{sleptsov2016vassiliev}. The resulting Vassiliev invariants up to 6-th order are presented in Appendix~\ref{sec:VI-pretzel(m,n,2)}.

 Using~\eqref{VI-2-dep}, one can prove that ${\cal V}_{2,1}^{P(m,n,\overline{2})}$ and ${\cal V}_{3,1}^{P(m,n,\overline{2})}$ are algebraically independent, while all ${\cal V}_{k,i}^{P(m,n,\overline{2})}$, $k\geq 4$, are algebraically dependent on ${\cal V}_{2,1}^{P(m,n,\overline{2})}$ and ${\cal V}_{3,1}^{P(m,n,\overline{2})}$. 
Vassiliev invariants ${\cal V}_{2,1}^{P(m,n,\overline{2})}$, ${\cal V}_{3,1}^{P(m,n,\overline{2})}$, ${\cal V}_{4,2}^{P(m,n,\overline{2})}$, ${\cal V}_{4,3}^{P(m,n,\overline{2})}$, ${\cal V}_{5,2}^{P(m,n,\overline{2})}$, ${\cal V}_{5,3}^{P(m,n,\overline{2})}$, ${\cal V}_{5,4}^{P(m,n,\overline{2})}$ are primary, while we hypothesize that Vassiliev invariants of higher orders are non-primary. For example, all ${\cal V}_{6,i}^{P(m,n,\overline{2})}$ can be expressed through other Vassiliev invariants of the same or lower order:

{\small \begin{equation}
\begin{aligned}
    0&=406 {\cal V}_{2,1}^3-542 {\cal V}_{2,1}^2-1080 {\cal V}_{3,1} {\cal V}_{2,1}-1020 {\cal V}_{4,2} {\cal V}_{2,1}-168 {\cal V}_{4,3} {\cal V}_{2,1}-123 {\cal V}_{2,1}-84 {\cal V}_{3,1}+732 {\cal V}_{4,2}-\\
    &-1872 {\cal V}_{4,3}+1212 {\cal V}_{5,2}-2508 {\cal V}_{5,3}-1140 {\cal V}_{5,4}+360 {\cal V}_{6,5}\,, \\
    0&=5 {\cal V}_{2,1}^3-105 {\cal V}_{2,1}^2-90 {\cal V}_{3,1} {\cal V}_{2,1}-45 {\cal V}_{4,2} {\cal V}_{2,1}+225 {\cal V}_{4,3} {\cal V}_{2,1}-92 {\cal V}_{2,1}+630 {\cal V}_{4,3}+30 {\cal V}_{5,2}+330 {\cal V}_{5,3}-300 {\cal V}_{5,4}+90 {\cal V}_{6,6}\,, \\
    0&=36 {\cal V}_{2,1}^3-92 {\cal V}_{2,1}^2-180 {\cal V}_{3,1} {\cal V}_{2,1}-30 {\cal V}_{4,2} {\cal V}_{2,1}-438 {\cal V}_{4,3} {\cal V}_{2,1}+29 {\cal V}_{2,1}-24 {\cal V}_{3,1}+222 {\cal V}_{4,2}-1002 {\cal V}_{4,3}+\\
    &+252 {\cal V}_{5,2}-828 {\cal V}_{5,3}+90 {\cal V}_{6,7}\,, \\
    0&=2 {\cal V}_{2,1}^3+6 {\cal V}_{2,1}^2-36 {\cal V}_{4,3} {\cal V}_{2,1}+65 {\cal V}_{2,1}+12 {\cal V}_{3,1}+84 {\cal V}_{4,2}-624 {\cal V}_{4,3}+84 {\cal V}_{5,2}-516 {\cal V}_{5,3}+60 {\cal V}_{5,4}+360 {\cal V}_{6,8}\,, \\
    0&=6 {\cal V}_{2,1}^3+18 {\cal V}_{2,1}^2-108 {\cal V}_{4,3} {\cal V}_{2,1}+19 {\cal V}_{2,1}-4 {\cal V}_{3,1}+12 {\cal V}_{4,2}-192 {\cal V}_{4,3}+12 {\cal V}_{5,2}-108 {\cal V}_{5,3}+60 {\cal V}_{5,4}+120 {\cal V}_{6,9}\,,
\end{aligned}
\end{equation}}

\noindent where we omit the superscript $P(m,n,\overline{2})$ for shortness. Again, either ${\cal V}_{2,1}^{P(m,n,\overline{2})}$ or ${\cal V}_{3,1}^{P(m,n,\overline{2})}$ is not complete knot invariant of $P(m,n,\overline{2})$ family, but both seem to be. We conjecture that the combination ${\cal V}_{3}^{P(m,n,\overline{2})}\left(\left({\cal V}_{2}^{P(m,n,\overline{2})}\right)^2+1\right)$ is complete knot invariant.

\subsection{Example of $P(m,m,\overline{2n})$ family}
Now, we consider pretzel knots $P(m,m,\overline{2n})$ with the parameter $m$ being odd. In this case, one cannot start with the fully symmetric function~\eqref{Vass-pretzel-gen} with the mirror symmetry condition~\eqref{pretzel-mirr} because two parameters of the family $P(m,m,\overline{2n})$ coincide. Instead, one starts with 
\begin{equation}
    {{\cal V}_{d}^{P(m,m,\overline{2n})} = \sum_{i=0}^d \sum_{j=0}^d a_{d,i,j} m^i n^j}
\end{equation}
and takes into account the topology of the family:
\begin{equation}
\begin{aligned}
    P(-m,-m,-\overline{2n}) = \overline{P(m,m,\overline{2n})}\,,\quad P(1,1,\overline{2n})=\overline{{\rm Tw}}_{2n}\,,
\end{aligned}
\end{equation}
and utilizes the obtained formulas for Vassiliev invariants for twist knots~\eqref{v-tw-gen}. However, already at the second order, the mentioned symmetries of pretzel knots are not enough to fix Vassiliev invariants entirely and one should use the known answers for Vassiliev invariants of some concrete pretzel knots~\cite{katlas} or just take formulas from~\cite{sleptsov2016vassiliev}. The resulting Vassiliev invariants up to 6-th order are listed in Appendix~\ref{sec:VI-pretzel(m,m,2n)}.

We use the criterium~\eqref{VI-2-dep}:
\begin{equation}
     \partial_m {\cal V}_{2,1}^{P(m,m,\overline{2n})}\cdot \partial_n {\cal V}_{3,1}^{P(m,m,\overline{2n})} - \partial_n {\cal V}_{2,1}^{P(m,m,\overline{2n})}\cdot \partial_m {\cal V}_{3,1}^{P(m,m,\overline{2n})} \neq 0
\end{equation}
to prove that ${\cal V}_{2,1}^{P(m,m,\overline{2n})}$ and ${\cal V}_{3,1}^{P(m,m,\overline{2n})}$ are algebraically independent, while all ${\cal V}_{k,i}^{P(m,m,\overline{2n})}$, $k\geq 4$, are algebraically dependent on ${\cal V}_{2,1}^{P(m,m,\overline{2n})}$ and ${\cal V}_{3,1}^{P(m,m,\overline{2n})}$. 
One can check that all Vassiliev invariants ${\cal V}_{2,1}$, ${\cal V}_{3,1}$, ${\cal V}_{4,2}$, ${\cal V}_{4,3}$, ${\cal V}_{5,2}$, ${\cal V}_{5,3}$, ${\cal V}_{5,4}$, ${\cal V}_{6,5}$, ${\cal V}_{6,6}$, ${\cal V}_{6,7}$, ${\cal V}_{6,8}$, ${\cal V}_{6,9}$ for $P(m,m,\overline{2n})$ are primary. We do not have an expression for the HOMFLY polynomial in the representation $[2,1]$ for this knot family; thus, we cannot calculate Vassiliev invariants of higher orders. So now, we cannot suppose whether this algebra of Vassiliev invariants is finitely or infinitely generated.

An interesting fact is that despite the fact that ${\cal V}_{2,1}^{P(m,m,\overline{2n})}$ and ${\cal V}_{3,1}^{P(m,m,\overline{2n})}$ are algebraically independent, they both do not form complete invariant of $P(m,m,\overline{2n})$ family. Instead, the set of ${\cal V}_{2,1}^{P(m,m,\overline{2n})}$, ${\cal V}_{4,2}^{P(m,m,\overline{2n})}$ seems to be complete knot invariant. We also suppose that the only Vassiliev invariant $({\cal V}_{2,1}^{P(m,m,\overline{2n})}+1){\cal V}_{4,2}^{P(m,m,\overline{2n})}$ is complete knot invariant.

\subsection{Towards a family with a single algebraically independent Vassiliev invariant}\label{sec:less-AI-VI}

In this section, we consider 2-parametric knot families which are candidates to have just a single algebraically independent Vassiliev invariant. We seek such knots among those possessing some symmetries. It is worth looking for 2-parametric knot families with (some) Vassiliev invariants dependent on less than two parameters. For example, amphichiral knots have zero Vassiliev invariants of all odd orders. We explore an amphichiral knot family in Section~\ref{sec:pretzel-v2=0}. One can also construct families of several first constant Vassiliev invariants using Stanford theorem~\cite{stanford1996braid,stanford1998vassiliev}, see an example in Section~\ref{sec:Stanford}. Another interesting knot family was given by Kanenobu~\cite{kanenobu1986examples,kanenobu1986infinitely}. Kanenobu knots form 2-parametric family with the fundamental HOMFLY polynomial dependent only on the sum of parameters. Thus, at least Vassiliev invariants of orders $\leq 4$ are polynomials only in one family parameter, see Section~\ref{sec:Kanenobu}. 

\subsubsection{Pretzel families with the third vanishing Vassiliev invariant}\label{sec:pretzel-v2=0}

We know the Vassiliev invariants for pretzel knots up to 6-th order explicitly~\cite{sleptsov2016vassiliev}. Thus, we find that the Vassiliev invariant ${\cal V}_{3,1}$ turns to zero for parallel pretzels $P(n,m,-n-m,\pm 1)$ and antiparallel ones $P(n,m,-n,-m)$.

\paragraph{Pretzel $P(n,m,-n-m,1)$} family includes amphichiral subfamily $P(n,1,-n-1,1)$. It includes the following knots from the Rolfsen table: $P(-2,1,1,1) = 4_1$, $P(2,1,-3,1) = 6_3$, $P(-4,3,1,1) = 8_9$, $P(2,3,-5,1) = 10_{48}$, $P(4,1,-5,1) = 10_{17}$, $P(4,3,-7,1) = 14a_{18462}$, $P(2,5,-7,1) = 14a_{18244}$. However, this family has two algebraically independent Vassiliev invariants:
\begin{equation}
\begin{aligned}
    {\cal V}_{2,1}^{P(n,m,-n-m,1)} &= (m+1) (m+n-1)\,, \\
    {\cal V}_{4,2}^{P(n,m,-n-m,1)} &= \frac{1}{12} (m+1) (m+n-1) \left(7 m^2+7 m n-4 n^2+5 n-1\right)\,.
\end{aligned}
\end{equation} 

\paragraph{$P(n,m,-n,-m)$} knots are amphichiral, and thus, having all ${\cal V}_{2k+1,i}\equiv 0$. However, this family has also two algebraically independent Vassiliev invariants:
\begin{equation}
\begin{aligned}
    {\cal V}_{2,1}^{P(n,m,-n,-m)} &= \frac{1}{2} \left(-2 m^2-2 n^2+3\right)\,, \\
    {\cal V}_{4,2}^{P(n,m,-n,-m)} &= \frac{1}{24} \left(16 m^4+24 m^2 n^2-44 m^2+16 n^4-44 n^2+33\right)\,.
\end{aligned}
\end{equation}

\subsubsection{Kanenobu family}\label{sec:Kanenobu}

Kanenobu knots $K(p,q)$ with $p=2n$ and $q=2m$ being numbers of crossings in the 2-strand braids, see Fig.\,\ref{fig:Kan-knots}, provide one of the first examples of the infinite number of distinct knots having the same HOMFLY polynomials in the fundamental representation. The fundamental HOMFLY polynomial depends only on the sum $p+q$. We have also calculated quantum knot invariants for the first symmetric representation and obtained the Vassiliev invariants up to the 6-th order:

\begin{equation}
\begin{aligned}
    d=2:\quad {\cal V}^{K(2n,2m)}_{2,1}&= -8\,, \\
    d=3:\quad {\cal V}^{K(2n,2m)}_{3,1}&= 16 (m+n)\,, \\
    d=4:\quad {\cal V}^{K(2n,2m)}_{4,1}&=\frac{1}{2}\left({\cal V}^{K(2n,2m)}_{2,1}\right)^2\,, \\
    {\cal V}^{K(2n,2m)}_{4,2}&= \frac{68}{3}-16 (m+n)^2\,, \\
    {\cal V}^{K(2n,2m)}_{4,3}&= \frac{28}{3}\,, \\
    d=5:\quad {\cal V}^{K(2n,2m)}_{5,1}&= {\cal V}^{K(2n,2m)}_{2,1} {\cal V}^{K(2n,2m)}_{3,1}\,, \\
    {\cal V}^{K(2n,2m)}_{5,2}&= \frac{32}{3} (m+n) \left((m+n)^2-14\right)\,, \\
    {\cal V}^{K(2n,2m)}_{5,3}&= -\frac{128}{3} (m+n)\,, \\
    {\cal V}^{K(2n,2m)}_{5,4}&= -16 (m+n)\,, \\
    d=6:\quad {\cal V}^{K(2n,2m)}_{6,1}&=\frac{1}{6}\left({\cal V}_{2,1}^{K(2n,2m)}\right)^3\,,\quad {\cal V}^{K(2n,2m)}_{6,2}=\frac{1}{2}\left({\cal V}^{K(2n,2m)}_{3,1}\right)^2\,,\\
    {\cal V}^{K(2n,2m)}_{6,3}&={\cal V}_{4,2}^{K(2n,2m)}{\cal V}_{2,1}^{K(2n,2m)}\,,\quad {\cal V}^{K(2n,2m)}_{6,4}={\cal V}_{4,3}^{K(2n,2m)}{\cal V}_{2,1}^{K(2n,2m)}\,, \\
    {\cal V}^{K(2n,2m)}_{6,5}&= -\frac{1}{3} 8 \left(2 (m+n)^2-143\right) (m+n)^2-\frac{1231}{15}\,, \\
    {\cal V}^{K(2n,2m)}_{6,6}&= \frac{4}{15} \left(40 \left(4 m^2-m n+4 n^2\right)+71\right)\,, \\
    {\cal V}^{K(2n,2m)}_{6,7}&= 16 \left(7 m^2+18 m n+7 n^2\right)-\frac{3484}{45}\,, \\
    {\cal V}^{K(2n,2m)}_{6,8}&= 8 \left(m^2+6 m n+n^2\right)+\frac{79}{9}\,, \\
    {\cal V}^{K(2n,2m)}_{6,9}&= 8 \left(m^2+6 m n+n^2\right)-\frac{271}{15}\,.
\end{aligned}
\end{equation}
We see that the Vassiliev invariants of orders $\leq 5$ depend only on the one parameter $n+m$. But at the 6-th order, there appear Vassiliev invariants dependent both on $n$ and $m$ separately. The Kanenobu family has two algebraically independent Vassiliev invariants ${\cal V}^{K(2n,2m)}_{3,1}$ and ${\cal V}^{K(2n,2m)}_{6,6}$.

\subsubsection{Stanford family}\label{sec:Stanford}

Let us construct the simplest Stanford 2-parametric knot family with the second vanishing Vassiliev invariant. We take the following 3-braids $p_{n,m}\in{\rm LCS}_3(P_3)$:
\begin{equation}
    p_{n,m}=[[\sigma_1^2,\sigma_2^{2n}],\sigma_1^{2m}]
\end{equation}
and $b=\sigma_1\sigma_2$, where $\sigma_1,\,\sigma_2\in {\cal B}_3$ are the braid group generators. Then, for the knot family ${\cal K}_{n,m}=\overline{p_{n,m}b}$ the first two Vassiliev invariants vanish: ${\cal V}_{2,1}^{{\cal K}_{n,m}}\equiv 0$ and ${\cal V}_{3,1}^{{\cal K}_{n,m}}\equiv 0$. However, one still can find two algebraically independent Vassiliev invariants:
\begin{equation}
\begin{aligned}
    {\cal V}_{4,2}^{{\cal K}_{n,m}} &= 8 m n (2 m-n+1)\,, \\
    {\cal V}_{5,2}^{{\cal K}_{n,m}} &= 8 m \left(m n^2-5 m n-n^2+5 n-4\right)\,.
\end{aligned}
\end{equation}

\setcounter{equation}{0}
\section{k-parametric families}

We now proceed to Vassiliev invariants of an arbitrary order $\leq d$ of a generic $k$-parametric family:
\begin{equation}\label{vk-gen}
    {\cal V}^{{\cal K}_{n_1,\dots,\,n_k}}_{d,i} = \sum\limits_{j_1,\dots,\,j_k=0}^d a_{d,\,j_1,\dots,\,j_k}^{(i)}\,\prod\limits_{i=1}^k n_i^{j_i}\,,\quad i = 1,\dots,\,\dim\left(\mathcal{A}_{d}\right)
\end{equation}
where $a_{d,\,j_1,\dots,\,j_k}^{(i)}\in \mathbb{Q}$ are some rational numbers. 

\paragraph{Primality.} The question on the number of generators of the algebra of such Vassiliev invariants is poorly studied. General argumentation tell that there may exist families with finitely generated algebras. However, there can appear topological symmetries that impose algebraic relations on Vassiliev invariants. It seems that this does not make 1-parametric Vassiliev invariants to be infinitely generated. However, in the case of $k>1$, relations on Vassiliev invariants can be restrictive enough for the algebra to be infinitely generated. An example is provided by the 2-parametric family of torus knots $T[m,n]$, see Section~\ref{sec:T[m,n]}.

\paragraph{Algebraic independence.} In a $k$-parametric knot family, there are $\leq k$ algebraically independent Vassiliev invariants. The algebraic independence of a set of Vassiliev invariants can be checked using the following theorem~\cite{lefschetz2005algebraic}. Polynomials $P_1$, \dots, $P_m$ in $x_1$, \dots, $x_n$ are algebraically independent iff the Jacobian matrix 
\begin{equation}
    \left(\frac{\partial P_i}{\partial x_j}\right)
\end{equation}
is of rank $m$, or equivalently for $m=n$ if and only if the Jacobian determinant is not zero. 

We have found only knot families with $k$ algebraically independent Vassiliev invariants. It is open question whether $k$-parametric knot families with $<k$ algebraically independent Vassiliev invariants exist. And if they do not exist, what is the reason? By now, nothing seems to contradict such an opportunity. 


\paragraph{Complete invariant.} As the number of Vassiliev invariants grows faster than the growth of monomials inside a Vassiliev invariant, then at some high enough order, all monomials in knot parameters are expressed through Vassiliev invariants. In particular, in a generic case, the parameters $n_1$, \dots, $n_k$ are expressed as linear combinations of Vassiliev invariants, being also Vassiliev invariants. Thus, this set of $k$ Vassiliev invariants becomes complete invariant of this knot family. Actually, a set of less than $k$ Vassiliev invariants could be a complete knot family invariant. But it is also open question to present such set explicitly.

\section{Conclusion}

In this paper, we have explored algebraic structures of correlators in the 3d Chern--Simons theory supplied with an arbitrary knot -- Vassiliev knot invariants. We have conducted the research focusing on $k$-parametric knot families with Vassiliev invariants polynomial in knot parameters. 

In particular, we pose question if the algebra of Vassiliev invariants is finitely or infinitely generated. It turns out that both cases are possible. Namely, the examples of finitely generated algebras are provided by torus $T[2,2k+1]$ and twist ${\rm Tw}_{2k}$ knots. We also conjecture that all algebras of 1-parametric polynomial Vassiliev invariants are finitely generated. Example of infinitely generated algebra is provided by torus knots $T[m,n]$.

We have also discovered that algebras of polynomial Vassiliev invariants possess additional algebraic relations. Then, the question is how many algebraically independent Vassiliev invariants such algebras have. The answer is that $k$-parametric algebra of Vassiliev invariants has $\leq k$ algebraically independent Vassiliev invariants. In fact, we have found only knot families having $k$ algebraically independent Vassiliev invariants, but not fewer.

The third problem we study concerns finding complete knot invariants. There is hypothesis that the set of all Vassiliev invariants is complete knot invariant. However, in practice, it cannot be used to distinguish all knots because the full algebra of Vassiliev invariants is infinitely generated and does not contain any other relations. Alternative is to split all knots into knot families and distinguish knots inside knot families. We have found out that for some $k$-parametric families with polynomial Vassiliev invariants, a set of $k$ Vassiliev invariants is complete invariant of this knot family what recovers an ability to distinguish knots by Vassiliev invariants.

\section*{Acknowledgements}

We are grateful for enlightening discussions to A. Morozov, And. Morozov, N. Tselousov, A. Popolitov and A. Bernakevich.

This work was supported in part by the Ministry of Science and Higher Education of the Russian Federation (agreement no. 075-03-2025-662). 


\printbibliography

\appendix

\setcounter{equation}{0}

\section{Vassiliev invariants for $P(m,n,\overline{2})$}\label{sec:VI-pretzel(m,n,2)}

Here we list the Vassiliev invariants for pretzel knots $P(m,n,\overline{2})$ up to the 6-th order:
\begin{equation}
\begin{aligned}
    d=2:\quad {\cal V}^{P(m,n,\overline{2})}_{2,1}&=\frac{1}{2} \left(m^2-4 m+n^2-4 n-2\right)\,, \\
    d=3:\quad {\cal V}^{P(m,n,\overline{2})}_{3,1}&=\frac{1}{3} (m+n-2) \left(m^2-m n-4 m+n^2-4 n-3\right)\,, \\
    d=4:\quad {\cal V}^{P(m,n,\overline{2})}_{4,1}&=\frac{1}{2}{\cal V}^2_{2,1}\,, \\ 
    {\cal V}^{P(m,n,\overline{2})}_{4,2}&=\frac{1}{24} \left(7 m^4-56 m^3-48 m^2 n+112 m^2-48 m n^2+192 m n+48 m+7 n^4-56 n^3+112 n^2+48 n-46\right)\,, \\
    {\cal V}^{P(m,n,\overline{2})}_{4,3}&=\frac{1}{24} \left(m^4-8 m^3+24 m^2+48 m n+16 m+n^4-8 n^3+24 n^2+16 n-2\right)\,, \\
    d=5:\quad {\cal V}^{P(m,n,\overline{2})}_{5,1}&={\cal V}_{2,1} {\cal V}_{3,1}\,, \\ 
    {\cal V}^{P(m,n,\overline{2})}_{5,2}&=\frac{1}{180} (51 m^5-510 m^4-420 m^3 n+1550 m^3-360 m^2 n^2+2700 m^2 n-600 m^2-\\
    &-420 m n^3+2700 m n^2-1800 m n-1661 m+51 n^5-510 n^4+1550 n^3-600 n^2-1661 n-60)\,, \\
    {\cal V}^{P(m,n,\overline{2})}_{5,3}&=\frac{1}{180} (9 m^5-90 m^4-60 m^3 n+290 m^3+540 m^2 n-120 m^2-60 m n^3+540 m n^2-\\ 
    &-360 m n-359 m+9 n^5-90 n^4+290 n^3-120 n^2-359 n-60)\,, \\
    {\cal V}^{P(m,n,\overline{2})}_{5,4}&=\frac{1}{30} \left(m^5-10 m^4+40 m^3+60 m^2 n-20 m^2+60 m n^2-60 m n-41 m+n^5-10 n^4+40 n^3-20 n^2-41 n\right) \\
    d=6:\quad {\cal V}^{P(m,n,\overline{2})}_{6,1}&=\frac{1}{6}{\cal V}^3_{2,1}\,,\quad {\cal V}^{P(m,n,\overline{2})}_{6,2}=\frac{1}{2}{\cal V}^2_{3,1}\,,\quad {\cal V}^{P(m,n,\overline{2})}_{6,3}={\cal V}_{4,2}{\cal V}_{2,1}\,,\quad {\cal V}^{P(m,n,\overline{2})}_{6,4}={\cal V}_{4,3}{\cal V}_{2,1}\,, \\ 
    {\cal V}^{P(m,n,\overline{2})}_{6,5}&= \frac{1}{720} (203 m^6-2436 m^5-2040 m^4 n+9800 m^4-1680 m^3 n^2+17040 m^3 n-\\
   &-11040 m^3-1680 m^2 n^3+18720 m^2 n^2-29760 m^2 n-9460 m^2-2040 m n^4+\\
   &+17040 m n^3-29760 m n^2-11040 m n+9144 m+203 n^6-2436 n^5+9800 n^4-\\
   &-11040 n^3-9460 n^2+9144 n+3234)\,, \\
   {\cal V}^{P(m,n,\overline{2})}_{6,6}&=\frac{1}{360} (5 m^6-60 m^5-180 m^4 n+90 m^4-120 m^3 n^2+660 m^3 n+220 m^3-120 m^2 n^3+\\
   &+900 m^2 n^2+420 m^2 n-101 m^2-180 m n^4+660 m n^3+420 m n^2-120 m n+224 m+\\
   &+5 n^6-60 n^5+90 n^4+220 n^3-101 n^2+224 n+312)\,, \\
   {\cal V}^{P(m,n,\overline{2})}_{6,7}&=\frac{1}{180} (18 m^6-216 m^5-60 m^4 n+1010 m^4+1380 m^3 n-1420 m^3+1260 m^2 n^2-\\
   &-3900 m^2 n-1087 m^2-60 m n^4+1380 m n^3-3900 m n^2-1440 m n+792 m+18 n^6-\\
   &-216 n^5+1010 n^4-1420 n^3-1087 n^2+792 n+178)\,, \\
   {\cal V}^{P(m,n,\overline{2})}_{6,8}&=\frac{1}{720} (m^6-12 m^5+60 m^4+120 m^3 n-40 m^3+360 m^2 n^2+360 m^2 n+234 m^2+\\
   &+120 m n^3+360 m n^2+720 m n+312 m+n^6-12 n^5+60 n^4-40 n^3+234 n^2+312 n+10)\,, \\
   {\cal V}^{P(m,n,\overline{2})}_{6,9}&=\frac{1}{240} (3 m^6-36 m^5+180 m^4+200 m^3 n-280 m^3+120 m^2 n^2-840 m^2 n-242 m^2+\\
   &+200 m n^3-840 m n^2-400 m n+72 m+3 n^6-36 n^5+180 n^4-280 n^3-242 n^2+72 n-2)\,.
\end{aligned}
\end{equation}

\setcounter{equation}{0}

\section{Vassiliev invariants for $P(m,m,\overline{2n})$}\label{sec:VI-pretzel(m,m,2n)}

Here we list the Vassiliev invariants for pretzel knots $P(m,m,\overline{2n})$ up to the 6-th order:
\begin{equation}
\begin{aligned}
    d=2:\quad {\cal V}^{P(m,m,\overline{2n})}_{2,1}&=m^2-4 m n-1\,, \\
    d=3:\quad {\cal V}^{P(m,m,\overline{2n})}_{3,1}&=\frac{2}{3} \left(m^3-9 m^2 n+6 m n^2-m+3 n\right)\,, \\
    d=4:\quad {\cal V}^{P(m,m,\overline{2n})}_{4,1}&=\frac{1}{2}{\cal V}^2_{2,1}\,, \\ 
    {\cal V}^{P(m,m,\overline{2n})}_{4,2}&=\frac{1}{12} \left(7 m^4-104 m^3 n+216 m^2 n^2-8 m^2-32 m n^3+80 m n-24 n^2+1\right)\,, \\
    {\cal V}^{P(m,m,\overline{2n})}_{4,3}&=\frac{1}{12} \left(m^4-8 m^3 n+48 m^2 n^2+16 m n-1\right)\,, \\
    d=5:\quad {\cal V}^{P(m,m,\overline{2n})}_{5,1}&={\cal V}_{2,1}{\cal V}_{3,1}\,, \\ 
    {\cal V}^{P(m,m,\overline{2n})}_{5,2}&=\frac{1}{90} (51 m^5-1110 m^4 n+4320 m^3 n^2-70 m^3-2760 m^2 n^3+1260 m^2 n+120 m n^4-\\
    &-1800 m n^2+19 m+120 n^3-150 n)\,, \\
    {\cal V}^{P(m,m,\overline{2n})}_{5,3}&=\frac{1}{90} \left(9 m^5-150 m^4 n+840 m^3 n^2-10 m^3-480 m^2 n^3+180 m^2 n-360 m n^2+m-30 n\right)\,, \\
    {\cal V}^{P(m,m,\overline{2n})}_{5,4}&=\frac{1}{15} m \left(m^4-10 m^3 n+100 m^2 n^2-80 m n^3+30 m n-40 n^2-1\right)\,, \\
    d=6:\quad {\cal V}^{P(m,m,\overline{2n})}_{6,1}&=\frac{1}{6}{\cal V}^3_{2,1}\,,\quad {\cal V}^{P(m,m,\overline{2n})}_{6,2}=\frac{1}{2}{\cal V}^2_{3,1}\,,\quad {\cal V}^{P(m,m,\overline{2n})}_{6,3}={\cal V}_{4,2}{\cal V}_{2,1}\,,\quad {\cal V}^{P(m,m,\overline{2n})}_{6,4}={\cal V}_{4,3}{\cal V}_{2,1}\,, \\ 
    {\cal V}^{P(m,m,\overline{2n})}_{6,5}&=\frac{1}{360} (203 m^6-6156 m^5 n+36540 m^4 n^2-340 m^4-49760 m^3 n^3+8960 m^3 n+\\
    &+12720 m^2 n^4-27840 m^2 n^2+140 m^2-192 m n^5+11840 m n^3-2504 m n-240 n^4+1860 n^2-3)\,, \\
    {\cal V}^{P(m,m,\overline{2n})}_{6,6}&=\frac{1}{180} (5 m^6-360 m^5 n+1230 m^4 n^2-30 m^4+480 m^3 n^3+160 m^3 n-180 m^2 n^2+\\
    &+19 m^2+320 m n^3-96 m n+150 n^2+6)\,, \\
    {\cal V}^{P(m,m,\overline{2n})}_{6,7}&=\frac{1}{90} (18 m^6-276 m^5 n+3030 m^4 n^2-10 m^4-6000 m^3 n^3+680 m^3 n+\\
    &+1200 m^2 n^4-3000 m^2 n^2-7 m^2+960 m n^3-168 m n+90 n^2-1)\,, \\
    {\cal V}^{P(m,m,\overline{2n})}_{6,8}&=\frac{1}{360} (m^6-12 m^5 n+360 m^4 n^2+160 m^3 n^3+160 m^3 n+480 m^2 n^4+120 m^2 n^2-\\
    &-6 m^2+320 m n^3-8 m n+5)\,, \\
    {\cal V}^{P(m,m,\overline{2n})}_{6,9}&=\frac{1}{360} (9 m^6-108 m^5 n+1320 m^4 n^2-3680 m^3 n^3+320 m^3 n+480 m^2 n^4-\\
    &-1800 m^2 n^2-6 m^2+320 m n^3-104 m n-3)\,.
\end{aligned}
\end{equation}

\end{document}